\documentclass[11pt,a4paper]{article}
\def\SSPSubmission{1}
\usepackage[T1]{fontenc}
\usepackage[utf8]{inputenc}
\usepackage{lmodern}
\usepackage{amsmath,amssymb,amsthm,mathtools,bm}
\usepackage[a4paper,margin=2.3cm,headheight=14pt]{geometry}
\usepackage{booktabs,array,enumitem,microtype,xcolor,needspace}
\usepackage{hyperref}
\usepackage[backend=biber,style=alphabetic,maxbibnames=8]{biblatex}
\usepackage{bookmark}
\usepackage{fancyhdr}
\setlist{itemsep=0.15em,topsep=0.3em}
\hypersetup{colorlinks=true,linkcolor=blue!50!black,
 citecolor=green!35!black,urlcolor=blue!50!black,
 pdfauthor={Junyang Liu, Hang Wang, Houjin Wang, Haoran Xu},
 pdftitle={Approximate Synthetic Spectra of Quantum Systems Associated with Compact Lie Groups},
 pdfkeywords={approximate synthetic spectrum; compact Lie group; highest weight; coherent state; coadjoint orbit}}

\theoremstyle{plain}
\newtheorem{theorem}{Theorem}[section]
\newtheorem{lemma}[theorem]{Lemma}
\newtheorem{proposition}[theorem]{Proposition}
\newtheorem{corollary}[theorem]{Corollary}
\theoremstyle{definition}
\newtheorem{definition}[theorem]{Definition}
\newtheorem{assumption}[theorem]{Assumption}
\theoremstyle{remark}
\newtheorem{remark}[theorem]{Remark}
\newcommand{\R}{\mathbb R}
\newcommand{\C}{\mathbb C}
\newcommand{\Z}{\mathbb Z}

\newcommand{\Hsp}{\mathcal H}
\newcommand{\gC}{\mathfrak g_{\C}}
\newcommand{\Orb}{\mathcal O}
\newcommand{\sSp}{\operatorname{sSp}}
\newcommand{\cut}{\theta}
\newcommand{\Cuts}{\Theta}

\DeclareMathOperator{\Ad}{Ad}

\DeclareMathOperator{\Dom}{Dom}
\DeclareMathOperator{\Var}{Var}
\DeclareMathOperator{\sd}{sd}
\DeclareMathOperator{\Cov}{Cov}
\DeclareMathOperator{\spec}{spec}

\DeclareMathOperator{\dist}{dist}
\DeclareMathOperator{\Span}{span}
\DeclareMathOperator{\tr}{tr}
\DeclareMathOperator{\Hom}{Hom}
\DeclareMathOperator{\Wt}{Wt}
\DeclareMathOperator{\Gr}{Gr}
\newcommand{\RE}{\operatorname{Re}}
\newcommand{\tGr}{\widetilde{\Gr}}
\newif\ifssphandbook
\ifdefined\SSPSubmission\ssphandbookfalse\else\ssphandbooktrue\fi

\title{Approximate Synthetic Spectra of Quantum Systems\\ Associated with Compact Lie Groups}
\author{Junyang Liu \quad Hang Wang \quad Houjin Wang \quad Haoran Xu}
\date{}
\begin{document}
\maketitle
\begin{abstract}
How can noncommuting quantum observables retain the geometry of a classical
phase space? Motivated by Mumford's question on approximately macroscopically
unique (AMU) states and the work of Lin and Wang, we study synthetic spectra
of position--momentum and angular-momentum systems. Under the hypotheses of
their AMU theorems, nonempty synthetic spectra yield AMU states. For
sufficiently small $\hbar$, the position--momentum spectrum fills the plane,
while the angular-momentum spectrum contains any prescribed ball. The latter
calculation leads to compact Lie groups: we prove that highest-weight
coadjoint orbits lie in regional synthetic spectra under an explicit variance
bound. Function-space representations of $SU(n)$, $SO(n)$, and $Sp(n)$ provide
further examples.
\end{abstract}
\ifssphandbook\tableofcontents\fi

\section{Introduction}\label{sec:introduction}
Commuting self-adjoint observables admit a joint functional calculus, and their
simultaneous values therefore have a precise spectral meaning. Quantum
observables usually do not commute. David Mumford asked how a recognizable,
approximately classical world can nonetheless arise from quantum measurements
\cite{Mumford2023}. His question brings two mathematical issues into focus.
One may try to approximate almost commuting observables by commuting ones.
Alternatively, one may seek unit states in which each observable has small
standard deviation: approximately macroscopically unique (AMU) states.
Approximation by a commuting tuple can produce such states under suitable
hypotheses, but AMU states can exist without a commuting model.

The distinction has a history. Lin proved a dimension-independent commuting
approximation theorem for pairs of almost commuting self-adjoint matrices
\cite{Lin1997}. In infinite dimension, almost commutation alone does not
ensure such an approximation. Lin later gave a sufficient condition involving
the synthetic and essential synthetic spectra, and explained the Fredholm-index
obstruction for pairs \cite{Lin2026Measurements}. For Mumford's AMU question,
Lin established a different bounded result: sufficiently small commutators
of self-adjoint contractions yield, near points of an appropriate synthetic
spectrum, states whose joint expectations are close to those points and whose
individual standard deviations are small \cite{Lin2024}. Closeness of
expectations and small fluctuations are separate estimates; neither asserts
that the operators themselves commute.

Position and momentum make the unbounded issue concrete. On a common invariant
core in $L^2(\R)$ they satisfy $[X,P_\hbar]=i\hbar I$, while both operators are
unbounded. Lin and Wang extended the AMU analysis to unbounded observables
under domain and commutator hypotheses \cite{LinWang2025}. Their work gives AMU states for position and momentum as $|\hbar|$
decreases. It also proves a positive lower bound for approximation by commuting
self-adjoint pairs, measured by operator norms of the differences on the common
invariant domain, even as $|\hbar|\to0$. Its stronger results treat angular
momentum, whose full-space
commutators are unbounded. These results prompt a geometric question: which
classical sets can be seen directly in a synthetic spectrum of the quantum
observables?

Compact group representations offer a natural answer. A strongly continuous
unitary representation $\rho$ of a compact connected Lie group $G$ and
$\hbar>0$ produce self-adjoint generators
\[
 A_X=-i\hbar\,d\rho(X),\qquad X\in\mathfrak g.
\]
For an irreducible block of highest weight $\lambda$, the expectation vectors
of its highest-weight coherent states sweep out the coadjoint orbit
$\Orb_{\hbar\lambda}$. This orbit interpretation is part of the geometric
picture behind representation theory and quantization
\cite{Kirillov2004,Landsman1998,Rieffel2004}. Angular momentum is the
familiar three-dimensional instance: its spin coherent states have expectation
vectors on a sphere. Our aim is to detect these orbits through the ordered
spectral cutoffs of the generators, with an explicit resolution bound. On
function-space representations, the commutators of the full generators are
often unbounded; finite-dimensional irreducible blocks let us obtain the
inclusion without assuming a small norm for those commutators.

For a self-adjoint tuple $A=(A_1,\ldots,A_N)$, let
$\Cuts_{\xi,\eta}(A)$ be the ordered product of spectral cutoffs centered at
$\xi\in\R^N$ at resolution $\eta$. We retain a grid center when this product
has norm at least $1-\eta$, and form the union of radius-$\eta$ balls about
retained centers in a bounded region. The resulting regional approximate
synthetic spectrum $\sSp^\eta_{D,M}(A)$ is defined precisely in
Section~\ref{sec:definitions}. Its cutoffs use bounded functional calculus,
including when the generators themselves are unbounded. The ordered product
follows the synthetic-spectrum constructions of
\cite{Lin2024,LinWang2025}; the grid and regional truncation are specified
here. This setting reduces orbit inclusion to two quantitative tasks:
control the loss in a product of noncommuting positive contractions and locate
a nearby grid center inside the chosen region.

\paragraph*{Main result.}
Fix $\hbar>0$ and an $\Ad$-invariant inner product on $\mathfrak g$, put
$N=\dim\mathfrak g\ge1$, and choose an orthonormal basis
$X_1,\ldots,X_N$. Let $\rho_{\rm W}$ be the half-sum of positive roots.
For $0<\eta\le 1/2$, set
\[
 \delta_\eta=\frac{\eta^2}{32N},\qquad
 D_\eta=\frac{2\delta_\eta}{\sqrt N}\Z^N.
\]
Theorem~\ref{thm:main} proves that, for an irreducible block of highest weight
$\lambda$ and $M\ge1$, the conditions
\[
 \hbar\|\lambda\|\le M-\delta_\eta,\qquad
 \hbar^2(\rho_{\rm W},\lambda)\le\frac{\eta^3}{16(N+1)}
\]
imply
\[
 \Orb_{\hbar\lambda}\subseteq\sSp^\eta_{D_\eta,M}(A).
\]
The inclusion is one-sided: it identifies a geometrically meaningful subset
without asserting that every point of the regional spectrum lies near an
orbit.

The proof has two exact ingredients. The coherent-state expectation map is
equivariant and has image $\Orb_{\hbar\lambda}$. The sum of the variances of
the $N$ generators in each such state is
$2\hbar^2(\rho_{\rm W},\lambda)$. A spectral tail estimate and a bound for
ordered products turn this variance into the cutoff threshold at a nearby
grid center; the radius condition keeps that center inside the region. The
same argument extends to selected blocks of a direct sum, including
representations with infinite multiplicities. The product estimate in
Lemma~\ref{lem:product} is related to quantum union bounds
\cite{OskoueiManciniWilde2019,ODonnellVenkateswaran2021}.

We first work out Gaussian position--momentum states and spin coherent states.
Their expectations fill a phase plane and spheres, respectively, and their
variances show directly how cutoff qualification arises. We then prove the
compact-group theorem, its direct-sum and boundary-grid variants, and a
tensor-power result in which a nonzero orbit remains fixed while a sufficient
resolution tends to zero. Natural function-space actions of $SU(n)$, $SO(n)$,
and $Sp(n)$ supply examples of the corresponding coadjoint-orbit geometry.
Reverse geometric bounds and optimal resolution remain open.

\section{The Synthetic Spectrum: Definitions}\label{sec:definitions}
\subsection{Approximate joint eigenstates and AMU states}
We use the convention that Hilbert-space inner products are linear in the
second variable, with norm $\|v\|=\sqrt{\langle v,v\rangle}$.
Self-adjointness means $A=A^*$ with identical domains. Fix a self-adjoint
tuple $A=(A_1,\ldots,A_N)$, $N\ge1$.

\begin{definition}\label{def:expectation}
Put $D_A=\bigcap_j\Dom A_j$. For $\psi\in D_A$, $\|\psi\|=1$, set
\[
 m_{\psi,j}=\langle\psi,A_j\psi\rangle,\qquad
 \Var_\psi(A_j)=\|(A_j-m_{\psi,j})\psi\|^2,\qquad
 \sd_\psi(A_j)=\sqrt{\Var_\psi(A_j)},
\]
and $m_\psi=(m_{\psi,1},\ldots,m_{\psi,N})$, $V_\psi=\sum_j\Var_\psi(A_j)$.
For $q\in\R^N$ and $\varepsilon>0$, define
\begin{align*}
 \mathrm{JAE}_\varepsilon(q;A)
 &=\{\psi\in D_A:\|\psi\|=1,\
            \max_j\|(A_j-q_j)\psi\|<\varepsilon\},\\
 \mathrm{AMU}_\varepsilon(A)
 &=\{\psi\in D_A:\|\psi\|=1,\
            \max_j\sd_\psi(A_j)<\varepsilon\}.
\end{align*}
\end{definition}
AMU abbreviates approximately macroscopically unique in the sense of
\cite{Mumford2023,Lin2024,LinWang2025}; its threshold is a standard deviation. The elementary identity
\begin{equation}\label{eq:variance-offset}
 \sum_j\|(A_j-q_j)\psi\|^2=V_\psi+|m_\psi-q|^2
\end{equation}
gives $\mathrm{JAE}_\varepsilon(q;A)\subseteq\mathrm{AMU}_\varepsilon(A)$.
These expressions use quadratic forms on $\Dom A_j$.
The operator expectation of $A_j^2$ itself is defined on $\Dom A_j^2$.
For such a unit $\psi$, the real covariance matrix is
\[
 \Cov_\psi(j,k)=\RE\langle
 (A_j-m_{\psi,j})\psi,(A_k-m_{\psi,k})\psi\rangle.
\]
It is symmetric and positive semidefinite, with trace $V_\psi$.

\subsection{Cutoff functions, grids, and regional truncation}
For a self-adjoint $A$, the spectral theorem assigns a projection-valued measure
$E_A$ and the bounded Borel functional calculus
$f(A)=\int f(t)\,dE_A(t)$.
For $\mu_\psi(S)=\langle\psi,E_A(S)\psi\rangle$ one has
\[
 \langle\psi,f(A)\psi\rangle=\int f\,d\mu_\psi,\qquad
 \|f(A)\psi\|^2=\int |f|^2\,d\mu_\psi,\qquad
 \|f(A)\|\le\|f\|_\infty.
\]
Analytic details are recorded in Appendix~\ref{app:prelim}.

\begin{definition}\label{def:cutoff}
For $a\in\R$ and $0<\eta<1$, let
\[
 \cut_{a,\eta}(t)=
 \begin{cases}
  1,&|t-a|\le3\eta/4,\\
  4-4|t-a|/\eta,&3\eta/4<|t-a|<\eta,\\
  0,&|t-a|\ge\eta.
 \end{cases}
\]
For $\xi\in\R^N$, put
$P_j(\xi)=\cut_{\xi_j,\eta}(A_j)$ and
$\Cuts_{\xi,\eta}(A)=P_1(\xi)\cdots P_N(\xi)$ in this fixed order.
\end{definition}
Each $P_j$ is a positive contraction: $0\le P_j\le I$.
An orthogonal projection additionally satisfies $P^2=P$.

\begin{definition}\label{def:ssp}
Write $B(x,r)=\{y:|y-x|\le r\}$. A set $D\subseteq\R^N$ is locally finite
if $D\cap K$ is finite for every compact $K$.
For such a grid and $R>0$, set
\begin{align*}
 Q_\eta(A)&=\{\xi\in\R^N:\|\Cuts_{\xi,\eta}(A)\|\ge1-\eta\},\\
 \sSp^\eta_D(A)&=\bigcup_{\xi\in D\cap Q_\eta(A)}B(\xi,\eta),\\
 \sSp^\eta_{D,R}(A)&=
 \bigcup_{\substack{\xi\in D\cap Q_\eta(A)\\|\xi|\le R}}B(\xi,\eta).
\end{align*}
\end{definition}
The global union is locally finite and closed; the regional union is finite and
compact, contained in $B(0,R+\eta)$. A covering radius $\varepsilon$ on a set
$K$ means $(\forall x\in K)(\exists\xi\in D)\ |x-\xi|\le\varepsilon$.
We use $\dist(x,S)=\inf_{y\in S}|x-y|$ and
$S[r]=\{x:\dist(x,S)\le r\}$.

Our default grid is
\begin{equation}\label{eq:grid}
 \delta_\eta=\frac{\eta^2}{32N},\qquad
 D_\eta=\frac{2\delta_\eta}{\sqrt N}\Z^N.
\end{equation}
Coordinate rounding proves its covering radius is at most $\delta_\eta$.
For $M>0$, define the finite regional grid
\begin{equation}\label{eq:regional-grid}
 E_{\eta,M}=p_M\bigl(D_\eta\cap B(0,M+\delta_\eta)\bigr),\qquad
 p_M(x)=
 \begin{cases}x,&|x|\le M,\\Mx/|x|,&|x|>M.\end{cases}
\end{equation}
Projection onto a closed Euclidean ball is nonexpansive; hence this grid covers
$B(0,M)$ within $\delta_\eta$, with all centers in that ball.

\subsection{Coadjoint orbits and coherent states}
For a Lie group $G$ with real Lie algebra $\mathfrak g=T_eG$, the adjoint
action is
$\Ad(g)=d(x\mapsto gxg^{-1})_e$; its dual action on the real vector space
$\mathfrak g^*=\Hom_{\R}(\mathfrak g,\R)$ is
\[
 (\Ad^*(g)\ell)(X)=\ell(\Ad(g^{-1})X),\qquad
 \Orb_\ell=\{\Ad^*(g)\ell:g\in G\}.
\]
For any action on a set, $G.x=\{g.x:g\in G\}$ and
$G_x=\{g:g.x=x\}$ denote its orbit and stabilizer.
For a complex vector space $V$, $\mathbb P(V)=(V\setminus\{0\})/\C^\times$
is its ray space. For a unitary representation $\rho$ and a unit vector $v$,
the vectors $\rho(g)v$ form its coherent-state family, and
$M_v=G.[v]\subseteq\mathbb P(V)$ is the corresponding ray orbit.

\section{Cutoff Estimates}\label{sec:cutoff}
\begin{lemma}[Cutoff tail bound]\label{lem:tail}
For a self-adjoint $A$, a unit $\psi\in\Dom A$, and
$P=\cut_{a,\eta}(A)$,
\begin{equation}\label{eq:tail}
 \|(I-P)\psi\|^2\le\langle\psi,(I-P)\psi\rangle
       \le\frac{16}{9\eta^2}\|(A-a)\psi\|^2.
\end{equation}
If $m=\langle\psi,A\psi\rangle$ and $|a-m|\le\eta/2$, then also
$\langle\psi,(I-P)\psi\rangle\le16\Var_\psi(A)/\eta^2$.
\end{lemma}
\begin{proof}
Pointwise,
$(1-\cut)^2\le1-\cut\le\mathbf1_{\{|t-a|>3\eta/4\}}
\le16(t-a)^2/(9\eta^2)$.
Integrate against the scalar spectral measure.
For the last bound, $|t-a|>3\eta/4$ implies $|t-m|>\eta/4$.
\end{proof}

\begin{lemma}[Ordered products of positive contractions]\label{lem:product}
For $0\le P_j\le I$ and a unit $\psi$, set
$a_j=\langle\psi,(I-P_j)\psi\rangle$.
Then
\begin{align}
 \|P_1\cdots P_N\|&\ge1-\sum_j\|(I-P_j)\psi\|,\label{eq:linear-product}\\
 \|P_1\cdots P_N\|&\ge1-\frac{N+1}{2}\sum_j a_j.\label{eq:product}
\end{align}
\end{lemma}
\begin{proof}
Put $Q_j=I-P_j$, $d_j=Q_j\psi$, $R_{j-1}=P_1\cdots P_{j-1}$.
Then $\|d_j\|^2\le a_j$ and
$I-P_1\cdots P_N=\sum_jR_{j-1}Q_j$.
Taking norms after applying to $\psi$ proves \eqref{eq:linear-product}.
The reversed telescoping expansion gives
$\|(R_{j-1}^*-I)\psi\|\le\sum_{i<j}\|d_i\|$. Hence
\begin{align*}
 1-\RE\langle\psi,P_1\cdots P_N\psi\rangle
 &\le\sum_j a_j+\sum_{i<j}\|d_i\|\|d_j\|\\
 &\le\sum_j a_j+\frac{N-1}{2}\sum_j\|d_j\|^2
 \le\frac{N+1}{2}\sum_j a_j.
\end{align*}
Finally $\|P_1\cdots P_N\|\ge\RE\langle\psi,P_1\cdots P_N\psi\rangle$.
\end{proof}

\begin{remark}[Comparison with a quantum union bound]\label{rem:union-bound}
Write $S=\sum_j a_j$. A projective union bound also yields
$\|P_1\cdots P_N\|\ge1-2S$.
Here is the conversion for the positive contractions used in this paper.
Adjoin a fresh two-dimensional factor for each $j$, and on its two components
put
\[
 \Pi_j=\begin{pmatrix}
 P_j&\sqrt{P_j(I-P_j)}\\
 \sqrt{P_j(I-P_j)}&I-P_j
 \end{pmatrix},\qquad
 R_j=I\otimes|0\rangle\langle0|_j .
\]
These are projections. On $\Omega=\psi\otimes|0\rangle^{\otimes N}$ their
individual failure probabilities are $a_j$ and $0$.
The successful sequence
$R_1\Pi_1\cdots R_N\Pi_N\Omega$
equals $(P_1\cdots P_N\psi)\otimes|0\rangle^{\otimes N}$.
For $S\le1$, the refined bound following Theorem 1.3 of
\cite{ODonnellVenkateswaran2021} gives failure at most $4S/(1+S)^2$.
Thus $\|P_1\cdots P_N\|\ge(1-S)/(1+S)\ge1-2S$; for $S>1$ the last
inequality is immediate.
This comparison improves the coefficient when $N\ge4$.
The elementary estimate \eqref{eq:product} is used for the main constants;
the next remark records the alternative supplied by this comparison.
\end{remark}

\begin{theorem}[Qualified-center criterion]\label{thm:criterion}
Let $A_1,\ldots,A_N$ be self-adjoint, $\psi\in D_A$ a unit vector.
If
\begin{equation}\label{eq:criterion}
 |\xi-m_\psi|\le\eta/2,\qquad V_\psi\le\frac{\eta^3}{8(N+1)},
\end{equation}
then $\xi\in Q_\eta(A)$. For arbitrary $\xi$ one also has
\begin{equation}\label{eq:offset}
 1-\|\Cuts_{\xi,\eta}(A)\|
 \le\frac{8(N+1)}{9\eta^2}\bigl(V_\psi+|m_\psi-\xi|^2\bigr).
\end{equation}
\end{theorem}
\begin{proof}
The last part of Lemma~\ref{lem:tail} gives
$\sum_j a_j\le16V_\psi/\eta^2$ under the displacement condition.
Lemma~\ref{lem:product} makes the loss at most
$8(N+1)V_\psi/\eta^2\le\eta$.
For \eqref{eq:offset}, instead sum \eqref{eq:tail} and use
\eqref{eq:variance-offset}.
\end{proof}

\begin{remark}[An alternative qualified-center bound]\label{rem:strong-criterion}
Under $|\xi-m_\psi|\le\eta/2$, Lemma~\ref{lem:tail} gives
$S\le16V_\psi/\eta^2$. The quantum union bound of
Remark~\ref{rem:union-bound} therefore yields
\[
 1-\|\Cuts_{\xi,\eta}(A)\|\le\frac{32V_\psi}{\eta^2}.
\]
In particular $V_\psi\le\eta^3/32$ also suffices for $\xi\in Q_\eta(A)$.
This improves the elementary variance condition for $N\ge4$, agrees with it
for $N=3$, and is more restrictive for $N=1,2$.
\end{remark}

\section{Two Model Systems: Position--Momentum and Angular Momentum}
\label{sec:models}
\subsection{The position--momentum system}
Fix $\hbar>0$. On $L^2(\R,dx)$ let
\[
 Xf(x)=xf(x),\quad \Dom X=\{f:xf\in L^2\},\qquad
 P_\hbar f=-i\hbar f',\quad \Dom P_\hbar=H^1(\R).
\]
These are self-adjoint, with common invariant core $\mathcal S(\R)$.
On that core $[X,P_\hbar]=i\hbar I$.
For any center $(q,p)$, the unit Gaussian
\[
 \psi_{q,p}(x)=(\pi\hbar)^{-1/4}
       e^{-(x-q)^2/(2\hbar)}e^{ipx/\hbar}
\]
has means $(q,p)$ and variances $\hbar/2,\hbar/2$.
The full domain and core argument, in arbitrary dimension, is given in
Appendix~\ref{app:heisenberg}.

\begin{proposition}\label{prop:position-momentum}
Let $M>0$ and $0<\eta\le1/2$. If either of the sufficient conditions
\[
 0<\hbar\le\eta^3/24
 \qquad\text{or}\qquad
 2e^{-9\eta^2/(32\hbar)}\le\eta
\]
holds, then
\[
 \sSp^\eta_{D_\eta}(X,P_\hbar)=\R^2,\qquad
 B(0,M)\subseteq\sSp^\eta_{E_{\eta,M},M}(X,P_\hbar)
       \subseteq B(0,M+\eta).
\]
\end{proposition}
\begin{proof}
In the first parameter range, use the Gaussian with mean equal to a prescribed
center. The cutoff tail bound
of Lemma~\ref{lem:tail} gives
$\sum_{j=1}^2\langle\psi,(I-P_j)\psi\rangle\le16\hbar/\eta^2$.
Lemma~\ref{lem:product} therefore gives
$\|\Cuts\|\ge1-24\hbar/\eta^2\ge1-\eta$.
For the second condition, the position density and the physical momentum
density of the centered Gaussian are both
$(\pi\hbar)^{-1/2}e^{-t^2/\hbar}$.
With $\operatorname{erfc}(s)=\frac2{\sqrt\pi}\int_s^\infty e^{-u^2}\,du$,
the scalar Gaussian tail inequality
$\operatorname{erfc}(s)\le e^{-s^2}$, $s\ge0$, gives
$\|(I-P_j)\psi\|\le e^{-9\eta^2/(32\hbar)}$.
For completeness, the tail inequality follows by differentiating
$e^{-s^2}-\operatorname{erfc}(s)$: it increases to $s=1/\sqrt\pi$ and then
decreases to zero, starting at zero.
The linear bound \eqref{eq:linear-product} now gives
$\|\Cuts\|\ge1-2e^{-9\eta^2/(32\hbar)}$.
Every center is qualified. The covering properties of the two grids give the
inclusions, and the regional definition gives the last containment.
\end{proof}

\subsection{The angular momentum system}
On $\mathcal S(\R^3)$ put
\[
 L_x=-i\hbar(y\partial_z-z\partial_y),\quad
 L_y=-i\hbar(z\partial_x-x\partial_z),\quad
 L_z=-i\hbar(x\partial_y-y\partial_x).
\]
Their self-adjoint closures generate rotations. Polar coordinates give
\begin{equation}\label{eq:angular-radial}
 L^2(\R^3)\simeq L^2((0,\infty),r^2\,dr)
       \widehat\otimes L^2(S^2,d\sigma),\qquad
 L^2(S^2)=\widehat\bigoplus_{l\ge0}\mathcal H_l.
\end{equation}
A fixed normalization accounts for the normalized area measure
$d\sigma$. Each angular block $\mathcal H_l$ has dimension $2l+1$ and
$\sum_jL_j^2=\hbar^2l(l+1)I$ on that block.
Thus the corresponding eigenspace in $L^2(\R^3)$ includes the full radial
factor. The decomposition and core assertions follow from the spherical
harmonic construction in Section~\ref{sec:so} and the direct sum facts in
Appendix~\ref{app:prelim}; see also \cite{sakurai2017,varshalovich1988,GoodmanWallach2009}.
More explicitly, finite angular sums with smooth compactly supported radial
factors away from zero lie in $\mathcal S(\R^3)$.
Angular truncation followed by approximation of the finitely many radial
coefficients gives graph-norm density for each generator.

Choose the normalized highest weight vector $|l,l\rangle$ in $\mathcal H_l$.
The usual ladder formulas give
\[
 \langle\vec L\rangle=(0,0,\hbar l),\qquad
 \Cov=\operatorname{diag}(\hbar^2l/2,\hbar^2l/2,0).
\]
For a rotation taking $e_3$ to $n\in S^2$, let
$|l,n\rangle=U_n|l,l\rangle$.
Then $\langle\vec L\rangle=\hbar l n$ and the total variance is $\hbar^2l$.
The ray $[|l,n\rangle]$ is independent of the chosen rotation.
For every rotation $R$,
$U_R|l,n\rangle=e^{i\gamma(R,n)}|l,Rn\rangle$ for a suitable phase;
the induced action on rays is strictly equivariant.

\begin{proposition}\label{prop:angular}
For $M\ge1$, $0<\eta\le1/2$, and
$0<\hbar\le\eta^3/[32(M+1)]$, the angular momentum tuple on either
$L^2(S^2)$ or $L^2(\R^3)$ satisfies
\[
 B(0,M)\subseteq\sSp^\eta_{E_{\eta,M},M}(L_x,L_y,L_z).
\]
For the default grid the same target is contained in
$\sSp^\eta_{D_\eta,M+\delta_\eta}(L_x,L_y,L_z)$.
\end{proposition}
\begin{proof}
Fix $y=rn$, $0\le r\le M$, and first choose a center with
$|\xi-y|\le\delta_\eta$: use the covering property of $E_{\eta,M}$ or $D_\eta$.
Then $|\xi|\le M$ in the first case and $|\xi|\le M+\delta_\eta$ in the
second. Independently choose $l\ge0$ with $|\hbar l-r|\le\hbar/2$.
The state $|l,n\rangle$ has mean $m=\hbar l n$ and
\[
 V=\hbar^2l\le\hbar(M+\hbar/2)
       \le\hbar(M+1)\le\frac{\eta^3}{32}.
\]
Here $\hbar<1$, and
$|\xi-m|\le\delta_\eta+\hbar/2\le\eta/192+\eta/512<\eta/2$.
The criterion of Theorem~\ref{thm:criterion}, with $N=3$, makes $\xi$
qualified, and $|y-\xi|\le\delta_\eta<\eta$ proves both inclusions.
The choice $l=0$ uses the constant angular state and has zero variance,
so it requires no separate exclusion near the origin.
On $L^2(\R^3)$ tensor this state with any unit radial function.
The generators act only on the finite-dimensional angular factor, so this
tensor lies in every $\Dom L_j$ by the product one-parameter groups;
no radial derivative is required.
\end{proof}

\subsection{From these systems to compact Lie groups}
For position and momentum, the relevant group is the three-dimensional
Heisenberg group $H_1$. Its nonzero central coadjoint orbits are
$\{(a,b,c):a,b\in\R\}\subseteq\mathfrak h_1^*$ at fixed $c\ne0$.
For angular momentum, $SO(3)$ acts on its dual by rotations and the orbit
at spin $l$ is the sphere of radius $\hbar l$.
In both cases a family of unit vectors fills the classical orbit by its
expectations, and small total variance ensures qualification of nearby centers.
For compact groups, highest weight vectors provide this construction uniformly.

\section{Compact Lie Groups: Setting}\label{sec:setting}
\subsection{A quantum system associated with a compact Lie group}
\begin{enumerate}[label=(\roman*)]
\item Let $G$ be a compact connected Lie group, $\mathfrak g=T_eG$,
$N=\dim\mathfrak g\ge1$. Fix an $\Ad$-invariant real inner product and an
orthonormal basis $X_1,\ldots,X_N$. The induced dual norm identifies
$\mathfrak g^*$ isometrically with $\R^N$.
\item Let $\Hsp\ne0$ be a separable complex Hilbert space and
$\rho:G\to U(\Hsp)$ a strongly continuous unitary representation; that is,
$g\mapsto\rho(g)v$ is continuous for every $v$.
\item For $X\in\mathfrak g$, Stone's theorem gives $\rho(\exp tX)=e^{itB_X}$ with self-adjoint
generator $B_X=-i\,d\rho(X)$ on the natural derivative domain. Set
\begin{equation}\label{eq:operators}
 A_X=\hbar B_X,\qquad A_j=A_{X_j},\qquad \hbar>0.
\end{equation}
The smooth vectors $\Hsp^\infty=\{v:g\mapsto\rho(g)v\text{ is smooth}\}$
form a common invariant core (Proposition~\ref{prop:core}).
\item On $\Hsp^\infty$,
\begin{equation}\label{eq:commutation}
 [A_X,A_Y]=-i\hbar A_{[X,Y]}.
\end{equation}
For unit smooth vectors, $\Cov_\psi(X,Y)$ denotes the real covariance
of $A_X,A_Y$.
\item Fix a maximal torus with algebra $\mathfrak t$ and a positive root
system $\Phi^+$. Here
$\mathfrak g_\alpha=\{Z\in\gC:[H,Z]=i\alpha(H)Z\ \forall H\in\mathfrak t\}$.
For $\mu\in\mathfrak t^*$ define $\mu^\sharp$ by
$(\mu^\sharp,H)=\mu(H)$, and
$(\mu,\nu)=(\mu^\sharp,\nu^\sharp)$.
Extend $\mu$ by zero on $\mathfrak t^\perp$ and put
$\rho_{\rm W}=\frac12\sum_{\alpha\in\Phi^+}\alpha$.
\end{enumerate}

\subsection{Structural assumptions and function-space realizations}
\begin{assumption}\label{asm:blocks}
Fix an orthogonal decomposition
$\Hsp=\widehat\bigoplus_{\nu\in\Lambda}V_\nu$ into nonzero finite-dimensional
irreducible $G$-invariant subspaces, with $\Lambda$ countable.
Each copy is counted separately.
\end{assumption}
The Peter--Weyl theorem guarantees this decomposition under the preceding hypotheses;
the standard statement and a proof of the direct sum domain formula are
recorded in Appendix~\ref{app:prelim}.
For each block the weight space and its support are
\[
 (V_\nu)_\mu=\{v\in V_\nu:B_Hv=\mu(H)v\ \forall H\in\mathfrak t\},
 \qquad \Wt(V_\nu)=\{\mu:(V_\nu)_\mu\ne0\}.
\]

\begin{assumption}\label{asm:highest}
Choose a block $V_0$, its highest weight $\lambda$, and a unit vector
$v_\lambda\in V_0$ such that
\begin{equation}\label{eq:highest}
 A_Hv_\lambda=\hbar\lambda(H)v_\lambda,\qquad
 d\rho(E_\alpha)v_\lambda=0\quad(\alpha\in\Phi^+).
\end{equation}
Here $E_\alpha$ is a unit root vector as normalized in
Appendix~\ref{app:root-planes}.
\end{assumption}
The highest weight theorem supplies this choice in every block.
All estimates are made in a chosen block. The blocks simultaneously reduce
every $A_j$, and the direct sum norm identity transfers the cutoff bounds
to the whole representation (Lemma~\ref{lem:block-calculus}).

\begin{proposition}\label{prop:l2-setting}
Let $X$ be a second-countable locally compact Hausdorff space with a Radon
measure $\mu$, and let $G$ act continuously preserving $\mu$.
Then $\rho(g)f=f\circ g^{-1}$ is a strongly continuous unitary representation
on the separable space $L^2(X,\mu)$.
For a product measure space $(X,\mu)=(Y\times Z,\nu\otimes\sigma)$,
where $\nu,\sigma$ are $\sigma$-finite Radon measures on second-countable
locally compact Hausdorff spaces, assume that $G$ acts only on $Z$ preserving
$\sigma$, and that $L^2(Y,\nu)$ is nonzero and separable. Then
\begin{equation}\label{eq:radial-product}
 L^2(X,\mu)=L^2(Y,\nu)\widehat\otimes L^2(Z,\sigma),\qquad
 \rho=I\otimes\rho_Z.
\end{equation}
The ordered cutoff norm equals the corresponding norm on $L^2(Z)$.
\end{proposition}
\begin{proof}
Continuous compactly supported functions are dense in $L^2(X,\mu)$.
On such functions continuity of the action and compactness of a sufficiently
small group neighborhood give a common compact support and convergence in
$L^2$; unitarity extends this to all vectors. Second countability and the
Radon property give separability. Fubini's theorem gives the tensor
identification, and the spectral calculus gives
$f(A_j)=I\otimes f(A_j^{(Z)})$.
Finally $\|I\otimes T\|=\|T\|$ for a nonzero Hilbert factor.
\end{proof}
For compact homogeneous $X=G/K$, pullback to $G$ selects the right $K$-fixed
part of $V_\rho\otimes V_\rho^*$ in Peter--Weyl, so the left block multiplicity
is $\dim(V_\rho^*)^K=\dim V_\rho^K$.
For a product $Y\times G/K$ with product measure and trivial action on $Y$,
an orthonormal basis of $L^2(Y)$ multiplies each angular multiplicity by
$\dim L^2(Y)$. In particular, a radial basis splits each angular block into
countably many finite-dimensional copies. Their generator domains and cutoff
norms follow from \eqref{eq:radial-product} and Appendix~\ref{app:direct-sums}.

\begin{remark}\label{rem:components}
For a compact Lie group with finitely many components, apply the connected
theory to the identity component $G_0$, whose Lie algebra is $\mathfrak g$.
The whole group permutes the resulting orbit types. Ray stabilizers in Proposition~\ref{prop:stabilizer} refer to the connected action on the selected $G_0$ block.
For $\dim G=0$, the dual is $\{0\}$ and the empty cutoff product on a nonzero
space is $I$, giving the spectrum $\{0\}$.
\end{remark}

\clearpage
\section{Main Result and Proof}\label{sec:main}
We combine the covariance at a highest weight vector with equivariance and
spectral tail estimates. The constants come from the fixed ordered product estimate of Lemma~\ref{lem:product}.

\subsection{The main theorem}
\begin{theorem}\label{thm:main}
Let $G$, $\rho$, and $\hbar>0$ be as in Section~\ref{sec:setting}.
Write $N=\dim\mathfrak g$, choose an orthonormal basis
$X_1,\ldots,X_N$, and set $A=(A_1,\ldots,A_N)$ with
$A_j=-i\hbar\,d\rho(X_j)$. Let $V_0$ be an irreducible summand of highest
weight $\lambda$, and let $\rho_{\rm W}$ denote the half-sum of the positive
roots. Fix $M\ge1$ and $0<\eta\le1/2$, and set
\[
 \delta_\eta=\frac{\eta^2}{32N},\qquad
 D_\eta=\frac{2\delta_\eta}{\sqrt N}\Z^N.
\]
If
\begin{equation}\label{eq:threshold}
 \hbar\|\lambda\|\le M-\delta_\eta,\qquad
 \hbar^2(\rho_{\rm W},\lambda)\le\frac{\eta^3}{16(N+1)},
\end{equation}
then the coadjoint orbit through $\hbar\lambda$ satisfies
\begin{equation}\label{eq:main}
 \Orb_{\hbar\lambda}\subseteq\sSp^\eta_{D_\eta,M}(A).
\end{equation}
The inclusion also holds for any union of block orbits each satisfying
\eqref{eq:threshold}. For an empty root system, $\rho_{\rm W}=0$.
\end{theorem}
\begin{remark}[Scope]\label{rem:scope}
The theorem gives a sufficient one-sided inclusion, with reverse distance
estimates and optimal thresholds left open.
\end{remark}

\subsection{Exact expectation and variance at a highest weight vector}
Use an orthonormal basis of $\mathfrak t$ and the real orthonormal root planes
$U_\alpha,V_\alpha$ normalized by
\[
 E_\alpha=(U_\alpha-iV_\alpha)/\sqrt2,\quad
 [H,U_\alpha]=\alpha(H)V_\alpha,\quad
 [H,V_\alpha]=-\alpha(H)U_\alpha,\quad
 [U_\alpha,V_\alpha]=\alpha^\sharp.
\]
Their construction is given in Appendix~\ref{app:root-planes}.

\begin{theorem}\label{thm:variance}
For $Y=H+\sum_{\alpha>0}(u_\alpha U_\alpha+v_\alpha V_\alpha)$,
\begin{equation}\label{eq:variance}
 m_{v_\lambda}(Y)=\hbar\lambda(Y),\qquad
 \Var_{v_\lambda}(A_Y)=
 \frac{\hbar^2}{2}\sum_{\alpha>0}(\lambda,\alpha)(u_\alpha^2+v_\alpha^2).
\end{equation}
Equivalently, as a real symmetric form,
\begin{equation}\label{eq:covariance}
 \Cov_{v_\lambda}
 =0|_{\mathfrak t}\ \oplus\
   \bigoplus_{\alpha>0}\frac{\hbar^2(\lambda,\alpha)}2 I_2.
\end{equation}
\end{theorem}
\begin{proof}
The Cartan directions have zero variance by \eqref{eq:highest}.
For one positive root, write $D_U=d\rho(U_\alpha)$,
$D_V=d\rho(V_\alpha)$ and $z=D_Vv_\lambda$.
The highest weight relation gives $D_Uv_\lambda=iz$.
Skew-adjointness in real directions and linearity in the second variable yield
\[
 \langle v_\lambda,D_UD_Vv_\lambda\rangle=i\|z\|^2,\qquad
 \langle v_\lambda,D_VD_Uv_\lambda\rangle=-i\|z\|^2.
\]
Subtracting gives
$2i\|z\|^2=\langle v_\lambda,d\rho(\alpha^\sharp)v_\lambda\rangle
=i(\lambda,\alpha)$.
Thus both real root directions have squared generator norm
$(\lambda,\alpha)/2$.

The root relation implies
$d\rho(\mathfrak g_\beta)(V_0)_\mu\subseteq(V_0)_{\mu+\beta}$:
commute $d\rho(H)$ past $d\rho(Z)$ and use $[H,Z]=i\beta(H)Z$.
Since the positive root part annihilates $v_\lambda$,
$D_Uv_\lambda,D_Vv_\lambda$ lie in the weight space
$(V_0)_{\lambda-\alpha}$.
They are orthogonal to $v_\lambda$, and the corresponding vectors for
different positive roots are orthogonal to one another.
For the same root, the two vectors differ by the factor $i$, so their real
covariance is zero. This proves \eqref{eq:covariance}, its associated quadratic
form, and the expectation formula.
\end{proof}

\begin{corollary}\label{cor:variance}
Set $C_\lambda=\frac12\max_{\alpha>0}(\lambda,\alpha)$ and
$L_G=\max_{\alpha\in\Phi}\|\alpha\|$, with both zero for empty $\Phi$.
Then
\begin{equation}\label{eq:variance-bound}
 \Var_{v_\lambda}(A_Y)\le\hbar^2C_\lambda\|Y\|^2,\qquad
 0\le C_\lambda\le\tfrac12 L_G\|\lambda\|,
\end{equation}
and for any orthonormal basis of $\mathfrak g$,
\begin{equation}\label{eq:total-variance}
 \sum_j\Var_{v_\lambda}(A_j)
   =\hbar^2\sum_{\alpha>0}(\lambda,\alpha)
   =2\hbar^2(\rho_{\rm W},\lambda).
\end{equation}
\end{corollary}
\begin{proof}
The first bound follows from the largest eigenvalue of the real covariance
form, and the second from Cauchy--Schwarz.
Its trace is independent of the chosen orthonormal basis and equals the sum
of the two equal entries in each root plane.
\end{proof}

\subsection{Transport and orbit identification}
\begin{proposition}\label{prop:transport}
Let $\psi_g=\rho(g)v_\lambda$. For real $X,Y\in\mathfrak g$,
\begin{align}
 m_{\psi_g}(X)&=(\Ad^*(g)\hbar\lambda)(X),\label{eq:transport-mean}\\
 \Var_{\psi_g}(A_Y)&=\Var_{v_\lambda}(A_{\Ad(g^{-1})Y}).\label{eq:transport-var}
\end{align}
Consequently the total variance is $2\hbar^2(\rho_{\rm W},\lambda)$ for every
$g$, and $\Var_{\psi_g}(A_Y)\le\hbar^2C_\lambda\|Y\|^2$.
\end{proposition}
\begin{proof}
Differentiating the representation's conjugation identity gives
$\rho(g)^{-1}A_X\rho(g)=A_{\Ad(g^{-1})X}$.
Take the expectation or the norm of the centered vector.
The adjoint action preserves the invariant inner product, so it preserves
the trace of covariance and the norm of $Y$.
\end{proof}

\begin{proposition}\label{prop:stabilizer}
Let $H_\lambda=\{g:\rho(g)v_\lambda\in\C^\times v_\lambda\}$ and
$G_\lambda=\{g:\Ad^*(g)\lambda=\lambda\}$.
Then $H_\lambda=G_\lambda$ and
\[
 \Orb_\lambda\longrightarrow M_\lambda=G.[v_\lambda],
 \qquad \Ad^*(g)\lambda\longmapsto[\rho(g)v_\lambda]
\]
is a diffeomorphism.
Moreover $\{m_{\psi_g}:g\in G\}=\Orb_{\hbar\lambda}$.
\end{proposition}
\begin{proof}
Transport gives $H_\lambda\subseteq G_\lambda$.
Choose a regular $H_0$ in the open chamber defining $\Phi^+$.
The highest weight theorem (Theorem~\ref{thm:highest-weight}) makes
$\lambda(H_0)$ the largest eigenvalue of $B_{H_0}$, with one-dimensional
eigenspace $\C v_\lambda$.
If $g\in G_\lambda$, transport gives
$\langle\rho(g)v_\lambda,B_{H_0}\rho(g)v_\lambda\rangle=\lambda(H_0)$.
Expanding in an eigenbasis shows that a unit vector attaining this expectation
belongs to that eigenspace. Thus $g\in H_\lambda$.
The closed subgroup theorem identifies both smooth orbits with the same
compact homogeneous space $G/G_\lambda$, proving the diffeomorphism
(cf.\ \cite[Proposition 4]{Landsman1998}).
The last assertion is \eqref{eq:transport-mean}.
\end{proof}

\begin{lemma}\label{lem:block-calculus}
For the block decomposition in Assumption~\ref{asm:blocks},
$f(A_j)=\bigoplus_\nu f(A_j|_{V_\nu})$ for bounded Borel $f$. Consequently
\[
 \|\Cuts_{\xi,\eta}(A)\|
   =\sup_\nu\|\Cuts_{\xi,\eta}(A|_{V_\nu})\|.
\]
\end{lemma}
\begin{proof}
A finite-dimensional invariant subspace for a self-adjoint generator is
reducing, as follows either from its spectral decomposition or the unitary
one-parameter group. The direct sum spectral calculus then gives both claims;
its natural domain is detailed in Appendix~\ref{app:direct-sums}.
\end{proof}

\subsection{Proof of the main theorem}
\begin{proof}[Proof of Theorem~\ref{thm:main}]
For $q\in\Orb_{\hbar\lambda}$ choose $\psi_g$ with mean $q$ by
Proposition~\ref{prop:stabilizer}. Invariance of the dual norm gives
$|q|=\hbar\|\lambda\|\le M-\delta_\eta$.
Choose $\xi\in D_\eta$ with $|\xi-q|\le\delta_\eta$.
Then $|\xi|\le M$, $|\xi-m_{\psi_g}|\le\delta_\eta\le\eta/2$, and
\[
 V_{\psi_g}=2\hbar^2(\rho_{\rm W},\lambda)
       \le\frac{\eta^3}{8(N+1)}.
\]
The qualified-center criterion, applied in the selected block, and
Lemma~\ref{lem:block-calculus} give $\xi\in Q_\eta(A)$.
Since $|q-\xi|\le\eta$, the ball $B(\xi,\eta)$ contains $q$.
Each orbit uses only its own block, so the inclusions can be united.
\end{proof}
Since $q=m_{\psi_g}$, applying \eqref{eq:variance-bound} to each
orthonormal direction and using Proposition~\ref{prop:transport} gives joint
approximate states at orbit points:
\begin{equation}\label{eq:joint-error}
 \max_j\|(A_j-q_j)\psi_g\|\le\hbar\sqrt{C_\lambda},
 \qquad q=m_{\psi_g}.
\end{equation}
In particular, $\psi_g\in\mathrm{AMU}_\varepsilon(A)$ for every
$\varepsilon>\hbar\sqrt{C_\lambda}$.

\begin{remark}[A strengthened orbit condition]\label{rem:strengthened}
With the same radius condition $\hbar\|\lambda\|\le M-\delta_\eta$, the
alternative sufficient variance condition
\begin{equation}\label{eq:strong-threshold}
 \hbar^2(\rho_{\rm W},\lambda)\le\frac{\eta^3}{64}
\end{equation}
also gives the inclusion in Theorem~\ref{thm:main}.
Indeed the coherent-state variance is $2\hbar^2(\rho_{\rm W},\lambda)$,
so Remark~\ref{rem:strong-criterion} applies to the same nearby centers.
For $N\ge4$ this is less restrictive than the variance condition in
\eqref{eq:threshold}; for $N=3$ they agree.
The direct sum and grid arguments apply with this alternative condition
as well. We keep the elementary constants in the three classical examples
to make their common derivation transparent. Neither version is claimed optimal.
\end{remark}

\subsection{Single blocks and direct sums}\label{sec:variants}
The proof applies to a single block, to a selected collection of blocks, and
to representations with repeated angular components.

\begin{theorem}[Direct sum version]\label{thm:direct-sum}
Let $\lambda_\nu$ be the highest weight of $V_\nu$.
For any subset $J\subseteq\Lambda$ whose members satisfy
\[
 \hbar\|\lambda_\nu\|\le M-\delta_\eta,\qquad
 \hbar^2(\rho_{\rm W},\lambda_\nu)\le\frac{\eta^3}{16(N+1)},
\]
one has
$\bigcup_{\nu\in J}\Orb_{\hbar\lambda_\nu}\subseteq\sSp^\eta_{D_\eta,M}(A)$.
A sufficient common variance condition on all radius-selected blocks is
\begin{equation}\label{eq:uniform}
 \hbar\|\rho_{\rm W}\|(M-\delta_\eta)
          \le\frac{\eta^3}{16(N+1)}.
\end{equation}
\end{theorem}
\begin{proof}
Use Theorem~\ref{thm:main} in each selected block.
Cauchy--Schwarz gives
$\hbar^2(\rho_{\rm W},\lambda_\nu)
\le\hbar\|\rho_{\rm W}\|\hbar\|\lambda_\nu\|$, proving the sufficient condition.
\end{proof}
The selected weights lie in a bounded part of the integral weight lattice,
so there are finitely many weight types at fixed $\hbar,M$.
Each type can have any countable multiplicity.
For example, on $L^2(S^2)$ one has $C_{l\varepsilon_1}=l/2$,
whereas for a torus $\rho_{\rm W}=0$ for every weight.
The relevant quantity in the variance selection is $(\rho_{\rm W},\lambda_\nu)$.

\subsection{Grid variants}
\begin{corollary}[Grid variants]\label{cor:grids}
In the main theorem and the direct sum version, replace $D_\eta$ by any
locally finite grid with covering radius $\varepsilon_\eta\le\eta/2$.
For centers restricted to $B(0,M)$ use the margin
$\hbar\|\lambda_\nu\|\le M-\varepsilon_\eta$.
Alternatively a target of radius at most $M$ gives centers in
$B(0,M+\varepsilon_\eta)$.
If a finite grid in $B(0,M)$ covers that ball within $\varepsilon_\eta$,
the target can have radius exactly $M$.
All three statements use the same variance bound as \eqref{eq:threshold}.
\end{corollary}
\begin{proof}
Choose a center within $\varepsilon_\eta$ of the coherent-state mean and apply
\eqref{eq:criterion}. The stated radius restrictions control its norm.
In particular, the grid $E_{\eta,M}$ includes boundary targets.
\end{proof}
If one instead chooses to use \eqref{eq:offset}, the additional inequality
$\varepsilon_\eta^2\le\eta^3/(N+1)$ is a sufficient budget at the same variance
bound. The direct criterion above already works for every
$\varepsilon_\eta\le\eta/2$.

\subsection{Fixed orbits along a tensor family}
\begin{proposition}[Fixed orbits along a tensor family]\label{prop:fixed-orbit}
Let $(\rho,V_0)$ be irreducible with highest weight $\lambda\neq0$, and for
each $k\ge1$ let
\[
 W_k=\Span\{\rho(g)^{\otimes k}v_\lambda^{\otimes k}:g\in G\}.
\]
Then $W_k$ is irreducible of highest weight $k\lambda$. Choose
\[
 M\ge\max\{1,\|\lambda\|\},\qquad
 K=\max\{1,16(N+1)M\|\rho_{\rm W}\|\},\qquad
 \hbar_k=1/k,\quad \eta_k=(K/k)^{1/3}.
\]
Define the unit states and the observables on $W_k$ by
\[
 \psi_{g,k}=\rho(g)^{\otimes k}v_\lambda^{\otimes k},\qquad
 A_j^{(k)}=-i\hbar_k\,d(\rho^{\otimes k})(X_j)\big|_{W_k}.
\]
For $k\ge\lceil8K\rceil$ the orbit does not move with $k$,
$\hbar_k\|k\lambda\|=\|\lambda\|\le M$, and
\begin{equation}\label{eq:fixed-orbit}
 \Orb_\lambda\subseteq
 \sSp^{\eta_k}_{E_{\eta_k,M},M}(A^{(k)}),
 \qquad\text{equivalently}\qquad
 \sup_{q\in\Orb_\lambda}\dist\bigl(q,\sSp^{\eta_k}_{E_{\eta_k,M},M}(A^{(k)})\bigr)=0.
\end{equation}
In addition, the test states satisfy, for every $q\in\Orb_\lambda$ and every $g$ with
$m_{\psi_{g,k}}=q$,
\begin{equation}\label{eq:fixed-orbit-joint}
 \max_j\|(A_j^{(k)}-q_j)\psi_{g,k}\|\le\sqrt{C_\lambda/k}.
\end{equation}
\end{proposition}
\begin{proof}
The span of the orbit of $v_\lambda^{\otimes k}$ is irreducible with extreme
weight $k\lambda$, by the standard highest weight tensor construction
\cite[Section~3]{Rieffel2004}.
On this block, $\hbar_k(k\lambda)=\lambda$, so the mean orbit of
Proposition~\ref{prop:stabilizer} is the fixed set $\Orb_\lambda$.
By \eqref{eq:total-variance} and $k\ge8K$,
\[
 V_{\psi_{g,k}}=2\hbar_k^2(\rho_{\rm W},k\lambda)
   =\frac{2(\rho_{\rm W},\lambda)}{k}
   \le\frac{2M\|\rho_{\rm W}\|}{k}
   \le\frac{K}{8(N+1)k}
   =\frac{\eta_k^3}{8(N+1)} .
\]
The first inequality is Cauchy--Schwarz with
$(\rho_{\rm W},\lambda)\le\|\rho_{\rm W}\|\|\lambda\|\le M\|\rho_{\rm W}\|$;
the second uses $K\ge16(N+1)M\|\rho_{\rm W}\|$.
We use the regional grid and the criterion directly, so the boundary
case $\|q\|=M$ requires no $M-\delta_{\eta_k}$ margin.
Fix $q\in\Orb_\lambda$ and choose $\xi\in E_{\eta_k,M}$ with
$|q-\xi|\le\delta_{\eta_k}$. Since $\delta_{\eta_k}\le\eta_k/2$, both
conditions in \eqref{eq:criterion} hold, and the qualified-center criterion
gives $\xi\in Q_{\eta_k}(A^{(k)})$. The projection defining
$E_{\eta_k,M}$ keeps $\xi$ in $B(0,M)$, so this center is regional, and
$q\in B(\xi,\eta_k)\subseteq\sSp^{\eta_k}_{E_{\eta_k,M},M}(A^{(k)})$.
Taking the supremum over $q$ proves the inclusion in \eqref{eq:fixed-orbit}.
For \eqref{eq:fixed-orbit-joint}, the constant of \eqref{eq:variance-bound} is
positively homogeneous along the ray through $\lambda$, so $C_{k\lambda}=kC_\lambda$, and
\eqref{eq:joint-error} in the block $W_k$ reads
$\hbar_k\sqrt{C_{k\lambda}}=\sqrt{C_\lambda/k}$.
\end{proof}
For this family the orbit is not shrunk to the origin, as it is when
$\hbar\to0$ at fixed representation, because the representation grows with $k$.
The two rates are distinct: $k^{-1/3}$ is a sufficient cutoff resolution for
the norm threshold, while $k^{-1/2}$ controls the joint residuals of the
test states.

Using Remark~\ref{rem:strengthened}, Proposition~\ref{prop:fixed-orbit}
also holds with $K=\max\{1,64M\|\rho_{\rm W}\|\}$ and the same definitions
of $\eta_k$ and $\hbar_k$: the variance is then at most $\eta_k^3/32$.
This alternative changes the sufficient constant, not the exponent.

\begin{remark}[Why the exponent is $1/3$ and $\hbar_k=1/k$]\label{rem:rate-scope}
Write $c=(\rho_{\rm W},\lambda)$. If $c>0$, the blockwise variance condition in
\eqref{eq:threshold} with $\hbar_k=k^{-\alpha}$ reads
\[
 \hbar_k^2(\rho_{\rm W},k\lambda)=c\,k^{1-2\alpha}
 \le\frac{\eta_k^3}{16(N+1)},
 \qquad\text{so}\qquad
 \eta_k\ge[16(N+1)c]^{1/3}k^{(1-2\alpha)/3}.
\]
At $\alpha=1$ the right side is $[16(N+1)c/k]^{1/3}$, which decreases; taking
$K\ge16(N+1)c$ treats all weights with $\|\lambda\|\le M$ at once. If $c=0$,
as for a nonzero character of a torus, the variance condition holds for every
$\eta_k$ and no exponent is forced. The radius of the orbit is
$\hbar_k\|k\lambda\|=k^{1-\alpha}\|\lambda\|$: it is constant only at
$\alpha=1$, it diverges for $\alpha<1$, and for $\alpha>1$ it tends to zero,
so these choices do not keep the nonzero target orbit fixed.
Although $\alpha>1$ permits a faster decay of the cutoff scale allowed by
the variance criterion, the target orbit then shrinks to the origin.
We choose $\alpha=1$ to retain a fixed nonzero orbit.
These are sufficient conditions for inclusion, not necessary lower bounds
on the achievable resolution.
\end{remark}

\section{\texorpdfstring{Type A: $SU(n)$ Acting on $L^2(\C^n)$}
{Type A: SU(n) Acting on L2(Cn)}}\label{sec:su}
We start with the natural action on the full Euclidean space, describe its
angular and projective versions, and then identify the geometry of the
coadjoint orbits supplied by their highest weight blocks.

\subsection{The natural action on Euclidean space}
Let $n\ge2$ and
$SU(n)=\{U\in M_n(\C):U^*U=I,\det U=1\}$.
Its linear action $z\mapsto Uz$ preserves Lebesgue measure on
$\C^n\simeq\R^{2n}$ and therefore defines the unitary representation
\[
 \Hsp=L^2(\C^n,dz;\C),\qquad \rho(U)f(z)=f(U^{-1}z).
\]
This representation is strongly continuous by Proposition~\ref{prop:l2-setting}.
Here $\mathfrak{su}(n)=\{X:X^*=-X,\tr X=0\}$.
For $X\in\mathfrak{su}(n)$, its self-adjoint observable is the natural
generator $A_X=-i\hbar\,d\rho(X)$; on $C_c^\infty(\C^n)$ its differential is
$d\rho(X)f(z)=\left.\frac{d}{dt}\right|_{0}f(e^{-tX}z)$.
With $(X,Y)=-2\tr(XY)$, choose an orthonormal basis and take $N=n^2-1$.

The action preserves the radius. Polar coordinates give
\[
 L^2(\C^n)\simeq L^2((0,\infty),r^{2n-1}dr)
       \widehat\otimes L^2(S^{2n-1},d\sigma),
\]
up to the fixed normalization of area measure.
The radial factor carries the identity representation. Hence it supplies
multiplicity, while the angular factor determines the ordered cutoff norms.
Choosing a radial orthonormal basis splits every angular irreducible block
into finite-dimensional copies on the full space.

For $H=i\operatorname{diag}(a_1,\ldots,a_n)$ with $\sum_i a_i=0$, write
$\varepsilon_i(H)=a_i$, $\lambda_j=\varepsilon_1+\cdots+\varepsilon_j$.
The positive roots are $\varepsilon_i-\varepsilon_j$, $i<j$, and
\begin{equation}\label{eq:su-metric}
 (\varepsilon_i,\varepsilon_j)=\tfrac12(\delta_{ij}-1/n),\qquad
 (\rho_{\rm W},\lambda_1)=(n-1)/4.
\end{equation}
The compact group $SU(n)$ is connected; its sphere action is transitive
with stabilizer $SU(n-1)$.

\subsection{Sphere and projective versions}
Let $\mathcal H_{p,q}$ denote restrictions to the unit sphere of harmonic
polynomials of degree $p$ in $z$ and $q$ in $\bar z$.
The angular and projective decompositions are
\begin{align*}
 L^2(S^{2n-1})&=\widehat\bigoplus_{p,q\ge0}\mathcal H_{p,q},
 &\lambda_{p,q}&=q\lambda_1+p\lambda_{n-1},\\
 L^2(\mathbb{CP}^{n-1})&=\widehat\bigoplus_{k\ge0}\mathcal H_{k,k},
 &\lambda_{k,k}&=k(\varepsilon_1-\varepsilon_n).
\end{align*}
Here $\mathbb{CP}^{n-1}=S^{2n-1}/U(1)$ has the pushed-forward probability
measure; its functions pull back to functions fixed by the scalar circle.
Thus the projective version selects $p=q$.
Each displayed block is irreducible. For $n=2$ the pairs $p+q=l$ give
$l+1$ equivalent spin-$l/2$ copies; for $n\ge3$ the weights are distinct.
The proof, including the highest vector
$z_n^p\bar z_1^q/\|z_n^p\bar z_1^q\|_2$, is in
Appendix~\ref{app:classical-complex}.

\begin{corollary}\label{cor:su-sphere}
For $(p,q)\ne(0,0)$, put
\[
 R_{p,q}^2=\frac{(n-1)(p^2+q^2)+2pq}{2n},\qquad
 (\rho_{\rm W},\lambda_{p,q})=\frac{(p+q)(n-1)}4 .
\]
If $M\ge1$, $0<\eta\le1/2$, and
\[
 \hbar R_{p,q}\le M-\delta_\eta,\qquad
 \hbar^2\le\frac{\eta^3}{4n^2(p+q)(n-1)},
\]
then $\Orb_{\hbar\lambda_{p,q}}\subseteq\sSp^\eta_{D_\eta,M}(A)$
on the sphere and on $L^2(\C^n)$.
For the projective block $\mathcal H_{k,k}$, $k\ge1$, the conditions become
\begin{equation}\label{eq:su-projective}
 \hbar k\le M-\delta_\eta,\qquad
 \hbar^2\le\frac{\eta^3}{8n^2k(n-1)}.
\end{equation}
\end{corollary}
\begin{proof}
Compute the weight norm with \eqref{eq:su-metric} and use
Theorem~\ref{thm:main}, since $N+1=n^2$.
The radial tensor identity preserves cutoff norms.
\end{proof}
The families $\mathcal H_{0,k}$ and $\mathcal H_{k,k}$ provide fixed
fundamental-weight and highest-root orbits at resolution $O(k^{-1/3})$
by Proposition~\ref{prop:fixed-orbit}.

\subsection{Geometry of the coadjoint orbits}\label{subsec:su-geometry}
The invariant metric identifies $\mathfrak{su}(n)^*$ with traceless
skew-Hermitian matrices. Under this identification $\hbar\lambda_{p,q}$
is represented by $i\hbar K_{p,q}/2$, where
\[
 K_{p,q}=\operatorname{diag}(q,0,\ldots,0,-p)-\frac{q-p}{n}I .
\]
Thus the coadjoint orbit consists of the unitary conjugates of this matrix
and is determined by its eigenvalues. Its radius is $\hbar R_{p,q}$.
The centralizer of this representative consists of the matrices preserving
each of its eigenspaces, giving the following descriptions.

If exactly one of $p,q$ is positive, the eigenspace multiplicities are
$1,n-1$ and
\[
 \Orb_{\hbar\lambda_{p,q}}\cong
 SU(n)/S(U(1)\times U(n-1))\cong\mathbb{CP}^{n-1}.
\]
It is the space of complex lines in $\C^n$, of real dimension $2n-2$.
In particular the fundamental block $\mathcal H_{0,1}$ gives this geometry.
The finite-dimensional natural module gives the same fundamental orbit;
its normalization is checked in Appendix~\ref{subsec:su-fundamental}.

If $p,q>0$ and $n\ge3$, there are three eigenspaces of dimensions $1,n-2,1$,
so
\[
 \Orb_{\hbar\lambda_{p,q}}\cong
 SU(n)/S(U(1)\times U(n-2)\times U(1)).
\]
This is the flag manifold
$\{L\subset P\subset\C^n:\dim_\C L=1,\ \dim_\C P=n-1\}$,
of real dimension $4n-6$.
The two one-dimensional eigenspaces determine $L$ and $P^\perp$.
In particular the nonconstant projective blocks $p=q=k$ give these
highest-root flag orbits.

For $n=2$ every nonzero $\lambda_{p,q}$ gives a two-sphere of radius
$\hbar(p+q)/2$. The constant block $p=q=0$ gives the point orbit $\{0\}$.

\section{\texorpdfstring{Types B/D: $SO(n)$ Acting on $L^2(\R^n)$}
{Types B/D: SO(n) Acting on L2(Rn)}}\label{sec:so}
The natural rotation action on Euclidean space reduces to its angular
representation. Its sphere and projective blocks lead to oriented
two-plane orbits, described in the final subsection.

\subsection{The natural action on Euclidean space}
For $n\ge3$, let
$SO(n)=\{g\in M_n(\R):g^{\mathsf T}g=I,\det g=1\}$.
Rotations preserve Lebesgue measure and define
\[
 \Hsp=L^2(\R^n,dx;\C),\qquad \rho(g)f(x)=f(g^{-1}x).
\]
This is a strongly continuous unitary representation.
Its algebra is $\mathfrak{so}(n)=\{X:X^{\mathsf T}=-X\}$.
Use the invariant metric $(X,Y)=-\tfrac12\tr(XY)$.
The matrices $E_{ij}=e_{ij}-e_{ji}$, $i<j$, form an orthonormal basis, with
$N=\binom n2$, and on $C_c^\infty(\R^n)$
\[
 d\rho(E_{ij})=D_{ij}=x_i\partial_j-x_j\partial_i,\qquad
 A_{ij}=-i\hbar D_{ij}.
\]
We take their natural self-adjoint generators as in Section~\ref{sec:setting}.
The group is compact and connected, by the planar-rotation normal form.
Polar coordinates give
\[
 L^2(\R^n)\simeq L^2((0,\infty),r^{n-1}dr)
       \widehat\otimes L^2(S^{n-1},d\sigma).
\]
Rotations act only on the sphere factor; the radial space supplies
countably many copies of each angular block, with the same cutoff norms.

\subsection{Sphere and projective versions}
Let $\mathcal H_l$ be the restrictions of degree-$l$ homogeneous harmonic
complex polynomials. Then
\[
 L^2(S^{n-1})=\widehat\bigoplus_{l\ge0}\mathcal H_l,\qquad
 L^2(\mathbb{RP}^{n-1})=\widehat\bigoplus_{k\ge0}\mathcal H_{2k}.
\]
Each $\mathcal H_l$ is irreducible, with highest weight $l\varepsilon_1$,
where $\varepsilon_1$ is dual to the Cartan direction $E_{12}$.
The projective quotient $S^{n-1}/\{\pm1\}$ selects the even functions and
therefore the even degrees. Both spaces carry the invariant probability
measure. The decomposition and irreducibility proofs are given in
Appendix~\ref{app:classical-real}.

\begin{corollary}\label{cor:so-sphere}
Let $n\ge3$, $l\ge1$, and $N=\binom n2$. The block $\mathcal H_l$ has
\begin{equation}\label{eq:so-constants}
 \|l\varepsilon_1\|=l,\qquad
 (\rho_{\rm W},l\varepsilon_1)=l(n-2)/2,\qquad C_{l\varepsilon_1}=l/2.
\end{equation}
For $M\ge1$, $0<\eta\le1/2$, sufficient conditions for its orbit inclusion are
\[
 \hbar l\le M-\delta_\eta,\qquad
 \hbar^2\le\frac{\eta^3}{8l(n-2)(N+1)}.
\]
The conclusion holds on the sphere and on $L^2(\R^n)$; for the projective
version use $l=2k$, $k\ge1$.
\end{corollary}
\begin{proof}
Use the constants proved in Appendix~\ref{app:classical-real} and
Theorem~\ref{thm:main}. The radial tensor factor preserves cutoff norms.
\end{proof}
The families $\mathcal H_k$ and $\mathcal H_{2k}$ also give fixed orbits by
Proposition~\ref{prop:fixed-orbit}.

The complete sphere representation gives a useful regional inclusion.
Set
$C_n=\{ru\wedge v:r\ge0,\ |u|=|v|=1,\ u\perp v\}$ and
$C_{n,M}=C_n\cap B(0,M)$.
\begin{theorem}\label{thm:cone}
For $n\ge3$, $M\ge1$, $0<\eta\le1/2$, and
$0<\hbar\le\eta^3/[8nM(N+1)]$,
\[
 C_{n,M}\subseteq\sSp^\eta_{D_\eta,M+\delta_\eta}(A),\qquad
 C_{n,M}\subseteq\sSp^\eta_{E_{\eta,M},M}(A).
\]
For every $y\in C_{n,M}$, every center $\xi\in\R^N$ with
$|\xi-y|\le\delta_\eta$ has cutoff norm
at least $1-\eta/4$.
\end{theorem}
The proof chooses $l$ with $|\hbar l-r|\le\hbar/2$ and applies the offset
estimate; the constants are checked in Appendix~\ref{app:cone-estimate}.

\subsection{Geometry of the coadjoint orbits}\label{subsec:plucker}
Identify $\mathfrak{so}(n)^*$ isometrically with $\Lambda^2\R^n$ by the
basis dual to $E_{ij}$:
$q=\sum_{i<j}q_{ij}e_i\wedge e_j$ and $|q|^2=\sum_{i<j}q_{ij}^2$.
The highest vector $(x_1+ix_2)^l/\|(x_1+ix_2)^l\|_2$ has mean
$\hbar l\,e_1\wedge e_2$, so
\begin{equation}\label{eq:so-orbit}
 \Orb_{\hbar l\varepsilon_1}
 =\{\hbar l\,u\wedge v:|u|=|v|=1,\ u\perp v\}.
\end{equation}
For $l>0$, such a bivector encodes an oriented two-plane and its fixed
area scale $\hbar l$. Its stabilizer consists of a rotation of the plane
and an independent rotation of its perpendicular complement. Hence
\[
 \Orb_{\hbar l\varepsilon_1}\cong
 SO(n)/(SO(2)\times SO(n-2))=\tGr(2,n),\qquad
 \dim_\R\tGr(2,n)=2(n-2).
\]
All oriented orthonormal two-frames extend to an $SO(n)$ matrix, which
proves transitivity and this quotient description.
The projective blocks give the same orbit geometry at radii $2\hbar k$.

The cone of these orbits also has explicit equations:
\begin{proposition}\label{prop:plucker}
For $q\in\Lambda^2\R^n$,
\[
 q\in C_n\quad\Longleftrightarrow\quad q\wedge q=0
 \quad\Longleftrightarrow\quad
 q_{ij}q_{kl}-q_{ik}q_{jl}+q_{il}q_{jk}=0\quad(i<j<k<l).
\]
Adding $|q|=r$ selects the orbit of radius $r$.
\end{proposition}
The skew-matrix normal form proves sufficiency of these Pl\"ucker relations;
the proof and the general invariant-polynomial orbit separation theorem are
in Appendix~\ref{app:classical-orbit-equations}.
The components transform together under rotations, while
$\|q\wedge q\|^2=0$ is a scalar invariant equation.

For $n=3$ all bivectors are simple:
$\tGr(2,3)\cong S^2$ and $C_3\simeq\R^3$.
Thus Theorem~\ref{thm:cone} includes the full regional ball.
For $n=4$, write
$L=(q_{23},q_{31},q_{12})$, $B=(q_{14},q_{24},q_{34})$ and
$J_\pm=(L\pm B)/2$, with $q_{31}=-q_{13}$.
Then
\[
 q\wedge q=0\Longleftrightarrow L\cdot B=0,\qquad
 |q|=r,\ L\cdot B=0\Longleftrightarrow |J_+|=|J_-|=r/2.
\]
The inverse $L=J_++J_-$, $B=J_+-J_-$ identifies each nonzero orbit
with $S^2_{r/2}\times S^2_{r/2}$.

\section{\texorpdfstring{Type C: $Sp(n)$ Acting on $L^2(\mathbb H^n)$}
{Type C: Sp(n) Acting on L2(Hn)}}\label{sec:sp}
We use the quaternionic linear action on Euclidean space and its sphere
and projective versions. Their highest weights produce projective,
isotropic Grassmann, and isotropic flag orbits.

\subsection{The natural action on Euclidean space}
Let $Sp(n)$ be the quaternionic unitary group acting on columns
$q\in\mathbb H^n$ by $q\mapsto gq$.
This action is orthogonal on $\R^{4n}$, so it induces
\[
 \Hsp=L^2(\mathbb H^n,dq;\C),\qquad \rho(g)f(q)=f(g^{-1}q).
\]
The function values are complex; $\mathbb H^n$ specifies the underlying
Euclidean space. Proposition~\ref{prop:l2-setting} gives strong continuity,
and $A_X=-i\hbar\,d\rho(X)$ denotes the natural self-adjoint generator.
On $C_c^\infty(\mathbb H^n)$,
$d\rho(X)f(q)=\left.\frac{d}{dt}\right|_0f(e^{-tX}q)$.

In the complex matrix realization,
\[
 Sp(n)=\{U\in U(2n):U^{\mathsf T}JU=J\},\qquad
 J=\begin{pmatrix}0&I\\-I&0\end{pmatrix}.
\]
The group is compact and connected, with transitive sphere action and
stabilizer $Sp(n-1)$.
Use $(X,Y)=-2\tr(XY)$ and $N=n(2n+1)$.
For $H=\operatorname{diag}(ia_1,\ldots,ia_n,-ia_1,\ldots,-ia_n)$ write
$\varepsilon_j(H)=a_j$; then
\begin{equation}\label{eq:sp-metric}
 (\varepsilon_i,\varepsilon_j)=\tfrac14\delta_{ij},\qquad
 \rho_{\rm W}=\sum_{j=1}^n(n+1-j)\varepsilon_j.
\end{equation}
The positive roots are $\varepsilon_i\pm\varepsilon_j$, $i<j$, and
$2\varepsilon_i$.
Finally,
\[
 L^2(\mathbb H^n)\simeq L^2((0,\infty),r^{4n-1}dr)
       \widehat\otimes L^2(S^{4n-1},d\sigma).
\]
The first factor is a trivial multiplicity space; angular cutoff norms
therefore also hold on the full Euclidean space.

\subsection{Sphere and projective versions}
Assume first $n\ge2$.
The commuting right action is $q\mapsto qu^{-1}$, $u\in Sp(1)$.
On the degree-$l$ spherical harmonics in real dimension $4n$,
\begin{equation}\label{eq:quaternion-decomposition}
 \mathcal H_l=
 \bigoplus_{\substack{a\ge b\ge0\\a+b=l}}
       \mathcal V_{(a,b)}\otimes\operatorname{Sym}^{a-b}\C^2,\qquad
 L^2(S^{4n-1})=\widehat\bigoplus_{l\ge0}\mathcal H_l.
\end{equation}
Here $\mathcal V_{(a,b)}$ has $Sp(n)$ highest weight
$\lambda_{a,b}=a\varepsilon_1+b\varepsilon_2$.
As an $Sp(n)$ module it occurs $a-b+1$ times; an orthonormal basis of the
second factor selects the individual irreducible copies.
The proof and its low-degree checks are in
Appendix~\ref{app:classical-quaternionic}.
Taking right $Sp(1)$ fixed functions gives
\[
 L^2(\mathbb{HP}^{n-1})=\widehat\bigoplus_{k\ge0}\mathcal V_{(k,k)},
 \qquad \mathbb{HP}^{n-1}=S^{4n-1}/Sp(1).
\]
The quotient has the pushed-forward probability measure.

\begin{corollary}\label{cor:sp-sphere}
For $a\ge b\ge0$, $a+b>0$,
\[
 \|\lambda_{a,b}\|=\tfrac12\sqrt{a^2+b^2},\qquad
 (\rho_{\rm W},\lambda_{a,b})=\frac{na+(n-1)b}{4}.
\]
For $M\ge1$, $0<\eta\le1/2$, sufficient conditions for its orbit inclusion are
\[
 \frac{\hbar}{2}\sqrt{a^2+b^2}\le M-\delta_\eta,\qquad
 \hbar^2\le\frac{\eta^3}{4(na+(n-1)b)(2n^2+n+1)}.
\]
For the projective block $a=b=k\ge1$, these become
\[
 \frac{\hbar k}{\sqrt2}\le M-\delta_\eta,\qquad
 \hbar^2\le\frac{\eta^3}{4k(2n-1)(2n^2+n+1)}.
\]
\end{corollary}
\begin{proof}
Use \eqref{eq:sp-metric} in Theorem~\ref{thm:main};
$N+1=2n^2+n+1$.
\end{proof}
The growing families $(k,0)$ and $(k,k)$ give fixed orbits by
Proposition~\ref{prop:fixed-orbit}.
For $n=1$, $Sp(1)=SU(2)$ acts freely and transitively on $S^3$:
$L^2(S^3)=\widehat\bigoplus_{a\ge0}V_a\otimes\C^{a+1}$,
where $V_a$ has weight $a\varepsilon_1$.
The constants above hold with $b=0$, while $\mathbb{HP}^0$ gives the
one-dimensional trivial representation.

\subsection{Geometry of the coadjoint orbits}\label{subsec:sp-geometry}
In the metric identification with $\mathfrak{sp}(n)$,
$\hbar\lambda_{a,b}$ is represented by
\[
 \frac{i\hbar}{4}\operatorname{diag}(a,b,0,\ldots,0,-a,-b,0,\ldots,0),
\]
with the positive and negative lists each of length $n$.
Its orbit has radius $\hbar\sqrt{a^2+b^2}/2$.
The centralizer of this diagonal matrix has a unitary factor for each group
of equal positive entries and a compact symplectic factor on the zero
quaternionic coordinates. For $n\ge2$ the nonzero cases are
\begin{center}
\begin{tabular}{@{}lll@{}}
\toprule
Parameters & Orbit & Real dimension\\
\midrule
$a>0,\ b=0$ & $Sp(n)/(U(1)\times Sp(n-1))$ & $4n-2$\\
$a=b>0$ & $Sp(n)/(U(2)\times Sp(n-2))$ & $8n-10$\\
$a>b>0$ & $Sp(n)/(U(1)^2\times Sp(n-2))$ & $8n-8$\\
\bottomrule
\end{tabular}
\end{center}
Here $Sp(0)=\{1\}$.
These dimension formulas follow by subtracting the stabilizer dimension
from $n(2n+1)$.

The first orbit is $\mathbb{CP}^{2n-1}$, the space of complex lines in
$\C^{2n}$. Transitivity on quaternionic unit vectors gives transitivity
on those lines. In particular the fundamental block $(a,b)=(1,0)$ has this
geometry; its root-vector normalization is checked in
Appendix~\ref{app:sp-fundamental}.

For the second orbit define the complex isotropic Grassmannian
\[
 \operatorname{IGr}(2,2n)=
 \{P\subseteq\C^{2n}:\dim_\C P=2,\ \omega|_{P\times P}=0\},
 \qquad \omega(z,w)=z^{\mathsf T}Jw.
\]
An orthonormal isotropic two-frame extends to a symplectic unitary basis.
The stabilizer of its two-plane is $U(2)\times Sp(n-2)$, proving that this
Grassmannian is exactly the orbit for $a=b>0$.
Thus functions on $\mathbb{HP}^{n-1}$ produce the mean orbits
$\operatorname{IGr}(2,2n)$ through their nonconstant blocks $(k,k)$.
The quaternionic projective base and the complex coadjoint orbit have these
distinct roles in the construction.

For $a>b>0$ the orbit additionally encodes a complex line in that isotropic
two-plane:
\[
 \{L\subset P\subset\C^{2n}:\dim_\C L=1,\ \dim_\C P=2,\ \omega|_P=0\}.
\]
Choosing the line splits the $U(2)$ stabilizer into $U(1)^2$, giving the last
row of the table.
The zero weight gives a point; for $n=1$ the nonzero orbits reduce to
$Sp(1)/U(1)\cong\mathbb{CP}^1\cong S^2$, of radius $\hbar a/2$.

\par\bigskip
\noindent\emph{Further questions.} Determine when a representation's regional
synthetic spectrum admits a reverse distance estimate to a specified union of
coadjoint orbits, and how the best cutoff resolution depends on the
representation, grid, and order of the factors. One may also seek analogous
orbit-inclusion and reverse-estimate results beyond compact groups, where
finite-dimensional complete reducibility may fail and different coherent-state
constructions are required.

\section*{Acknowledgments and disclosure}
We thank Huaxin Lin for suggesting the position--momentum and
angular-momentum synthetic-spectrum calculations, which led us to further
examples and to a link between operator-algebraic methods and representation
theory. Hang Wang was supported by grant 23JC1401900 and National Natural
Science Foundation of China grant 12271165; she was also partially supported
by the Science and Technology Commission of Shanghai Municipality
(No.~22DZ2229014) and the Fundamental and Interdisciplinary Disciplines
Breakthrough Plan of the Ministry of Education of China (FIDBP). The other
three authors, all undergraduates, were supported by the Cultivation and
Advanced Projects of
East China Normal University's Outstanding Academic Training Program for
Future Leading Talents. AI tools assisted in preparing the type A--D example
calculations; the authors are responsible for the final mathematical
statements and proofs.

\appendix
\section{Background and Technical Details}\label{app:prelim}
We collect the analytic facts and representation-theoretic conventions used
in the proofs, followed by the technical details for the classical examples.
Throughout this appendix, $G$ is compact and connected, $\rho$ denotes a
representation, and $\rho_{\rm W}$ is the half-sum of positive roots.
Standard results are cited; the domain arguments and normalization checks
needed for our estimates are included explicitly.

\subsection{Analytic conventions and approximation}\label{app:analysis}
We use the inner product and adjoint conventions of
Section~\ref{sec:definitions}. For a densely defined operator $A$, the
graph norm is $\|v\|_A=(\|v\|^2+\|Av\|^2)^{1/2}$; a subspace of
$\Dom A$ is a core when it is dense in this norm. An operator is
essentially self-adjoint when its closure is self-adjoint.
The spectral theorem and functional calculus are used in their standard
forms; see \cite[Chapters~3--4]{Kowalski2009} and \cite{Rudin1991}.

\begin{lemma}\label{lem:finite-spectral}
A finite-dimensional self-adjoint operator has an orthonormal eigenbasis
with real eigenvalues; a commuting self-adjoint family has a joint such
basis. For a positive contraction $P$, $(I-P)^2\le I-P$ and $\|P\|\le1$.
\end{lemma}
The last two inequalities follow from $(1-p)^2\le1-p$ for $0\le p\le1$.
For self-adjoint $A$, write $E_A$ for its projection-valued spectral measure
and $\mu_\psi(S)=\langle\psi,E_A(S)\psi\rangle$. Then
\[
 A=\int t\,dE_A(t),\qquad
 \Dom A=\left\{\psi:\int t^2\,d\mu_\psi(t)<\infty\right\},
 \qquad
 \|(A-a)\psi\|^2=\int(t-a)^2\,d\mu_\psi(t).
\]
The last identity holds for $\psi\in\Dom A$ and is the spectral input to
the cutoff tail estimate. Our Fourier convention is
\[
 (\mathcal Ff)(\xi)=(2\pi)^{-d/2}\int_{\R^d}f(x)e^{-ix\cdot\xi}\,dx,
 \qquad
 \mathcal F(-i\hbar\partial_j)\mathcal F^{-1}=\hbar M_{\xi_j}.
\]
Plancherel extends $\mathcal F$ unitarily to $L^2$; the operator identity
uses the natural self-adjoint domains.

\begin{lemma}[Polynomial approximation]\label{lem:bernstein}
Every continuous real function on a closed Euclidean box is uniformly
approximable by real polynomials. The complex-valued statement follows by
approximating real and imaginary parts.
\end{lemma}
This is the standard polynomial approximation theorem on compact sets
\cite{Rudin1991}. Only this statement is needed for harmonic density and
invariant-polynomial separation below.

\subsection{Compact representations and smooth cores}
\label{app:haar}\label{app:peter-weyl}\label{app:stone}
Normalized Haar measure has mass one and is both left and right invariant.
Averaging any real inner product on $\mathfrak g$ gives an invariant one,
whose differentiation yields
\begin{equation}\label{eq:invariant-inner}
 ([X,Y],Z)=-(Y,[X,Z]).
\end{equation}
Consequently $(\Ad^*(g)\ell)^\sharp=\Ad(g)\ell^\sharp$.
Invariant orthogonal complements give finite-dimensional complete
reducibility; Schur orthogonality normalizes matrix coefficients.

\begin{lemma}\label{lem:smooth-representation}
A finite-dimensional continuous representation of a Lie group is smooth.
For a unitary representation on such a space,
\[
 d\rho(X)^*=-d\rho(X),\qquad
 d\rho([X,Y])=[d\rho(X),d\rho(Y)],\qquad
 \rho(g)^{-1}d\rho(X)\rho(g)=d\rho(\Ad(g^{-1})X).
\]
\end{lemma}
\begin{proof}
Convolution with a smooth approximate identity gives operators
$\rho(k_\varepsilon)$ whose images consist of smooth vectors.
They converge to $I$ in norm in finite dimension, so one is invertible;
its image is the whole space. Differentiating unitarity, commutators,
and conjugation gives the displayed identities.
\end{proof}

\begin{theorem}[Peter--Weyl]\label{thm:peter-weyl}
For the left regular action,
\[
 L^2(G)\simeq\widehat\bigoplus_{[\rho]\in\widehat G}
              V_\rho\otimes V_\rho^*.
\]
Every strongly continuous unitary representation is a Hilbert direct sum
of finite-dimensional irreducibles. On a separable space there are at
most countably many nonzero copies, counted with multiplicity.
\end{theorem}
We use this standard theorem in the form recorded in
\cite[Appendix Z.3]{ConradLandesman2018}.
Here $V_\rho^*$ is the complex linear dual. For $d_\rho=\dim V_\rho$,
the isometry
$u\otimes\ell\mapsto\sqrt{d_\rho}\,\ell(\rho(g^{-1})u)$
realizes the displayed summand; left translation acts on $u$ by $\rho$.
This fixes the convention used for homogeneous-space multiplicities.

By Stone's theorem \cite[Theorem~6.4]{Kowalski2009},
$U(t)=\rho(\exp tX)=e^{itB_X}$, where
\[
 \Dom B_X=\left\{v:\lim_{t\to0}\frac{U(t)v-v}{it}
                          \text{ exists in }\Hsp\right\},\qquad
 B_Xv=\lim_{t\to0}\frac{U(t)v-v}{it}.
\]
Thus $B_X=-i\,d\rho(X)$ on the natural derivative domain.

\begin{proposition}\label{prop:core}
For a strongly continuous unitary representation of $G$ on a separable
Hilbert space, the smooth
vectors $\Hsp^\infty$ are dense and a common invariant core for the
generators $B_X$. In particular their restrictions to $\Hsp^\infty$
are essentially self-adjoint.
\end{proposition}
\begin{proof}
In the separable setting, let $Q_m$ project onto the first $m$ blocks of
the Peter--Weyl decomposition. Their vectors are smooth by
Lemma~\ref{lem:smooth-representation}. The projections commute with
$U(t)$, so the derivative definition gives
$B_XQ_mv=Q_mB_Xv$ for $v\in\Dom B_X$.
Both $Q_mv\to v$ and $B_XQ_mv\to B_Xv$, proving graph-norm density.
The orbit map definition of smooth vectors shows invariance under
$\rho(G)$ and under each differentiated action.
\end{proof}

\subsection{Direct sums and functional calculus}\label{app:direct-sums}
For self-adjoint matrices $A_\nu$ on finite-dimensional blocks $V_\nu$,
their self-adjoint direct sum on $\widehat\bigoplus_\nu V_\nu$ is
\[
 \Dom A=\left\{(v_\nu):\sum_\nu\|A_\nu v_\nu\|^2<\infty\right\},
 \qquad A(v_\nu)=(A_\nu v_\nu).
\]
Testing the adjoint against single-block vectors gives exactly this
domain and action for $A^*$, hence $A=A^*$.
Finite-block truncations converge in graph norm.
For a compact representation,
$\rho(\exp tX)|_{V_\nu}=e^{itB_{X,\nu}}$ and
$\|(e^{itB_{X,\nu}}-I)v_\nu/t\|\le\|B_{X,\nu}v_\nu\|$.
Dominated convergence gives the strong derivative on the displayed
domain. Conversely, any strong derivative has these block components
and hence the same square summability condition. This identifies the
direct sum with the natural Stone generator, including its full domain.
Spectral measures and bounded Borel functions act blockwise; in particular,
\[
 \Cuts_{\xi,\eta}(A)=\bigoplus_\nu\Cuts_{\xi,\eta}(A^{(\nu)}),
 \qquad
 \|\Cuts_{\xi,\eta}(A)\|=\sup_\nu\|\Cuts_{\xi,\eta}(A^{(\nu)})\|.
\]

\subsection{Weights, roots, and real normalization}
\label{app:roots}\label{app:real-structure}\label{app:root-planes}
Let $T$ be a maximal torus, with algebra $\mathfrak t$.
Then $T=\mathfrak t/\Lambda_T$, where
$\Lambda_T=\ker(\exp|_{\mathfrak t})$ is a full-rank lattice, and
\[
 P_T=\{\mu\in\mathfrak t^*:\mu(\Lambda_T)\subseteq2\pi\Z\}
\]
is its integral weight lattice.
These structural facts and the root decomposition are standard
\cite[Chapters~2--3]{GoodmanWallach2009}; see also
\cite{Varadarajan1984} for compact Lie groups.
The algebra and its dual are real, while $\Hsp$ is complex.
On $\gC=\mathfrak g\oplus i\mathfrak g$ use compact conjugation
$\overline{X+iY}=X-iY$ and extend $d\rho$ complex linearly on smooth vectors.
Skew-symmetry in real directions gives
\begin{equation}\label{eq:complex-adjoint}
 d\rho(Z)^\dagger=-d\rho(\bar Z).
\end{equation}
Here $\dagger$ denotes a formal adjoint on the smooth domain and the
ordinary adjoint in a finite-dimensional block. For matrix compact forms,
$\bar Z=-Z^*$.

Let $b$ be the complex bilinear extension of the invariant real metric,
and $\langle Z,W\rangle=b(\bar Z,W)$ its Hermitian extension.
Our roots are real functionals with the convention
\begin{equation}\label{eq:roots}
 \gC=\mathfrak t_\C\oplus\bigoplus_{\alpha\in\Phi}\mathfrak g_\alpha,
 \qquad
 \mathfrak g_\alpha=\{Z:[H,Z]=i\alpha(H)Z\ \forall H\in\mathfrak t\}.
\end{equation}
Each root space is one-dimensional; distinct spaces are Hermitian
orthogonal, $\overline{\mathfrak g_\alpha}=\mathfrak g_{-\alpha}$, and
$[\mathfrak g_\alpha,\mathfrak g_\beta]\subseteq\mathfrak g_{\alpha+\beta}$,
with zero subscript interpreted as $\mathfrak t_\C$.

The normalization is relevant to every variance constant in the paper.
For a unit $Z\in\mathfrak g_\alpha$, invariance implies
$2i\alpha(H)b(Z,Z)=0$, hence $b(Z,Z)=0$.
Writing $Z=(U_\alpha-iV_\alpha)/\sqrt2$ therefore gives real
orthonormal $U_\alpha,V_\alpha$, and
\begin{equation}\label{eq:root-plane}
 [H,U_\alpha]=\alpha(H)V_\alpha,\quad
 [H,V_\alpha]=-\alpha(H)U_\alpha,\quad
 [U_\alpha,V_\alpha]=\alpha^\sharp,\quad
 [Z,\bar Z]=i\alpha^\sharp.
\end{equation}
Indeed $[Z,\bar Z]\in\mathfrak t_\C$ and
$([U_\alpha,V_\alpha],H)=(U_\alpha,[V_\alpha,H])=\alpha(H)$.
We take $E_\alpha=Z$, $E_{-\alpha}=\bar Z$, so that
$\mathfrak g=\mathfrak t\oplus\bigoplus_{\alpha>0}
(\R U_\alpha\oplus\R V_\alpha)$ orthogonally.

\subsection{Highest weights and the Casimir scalar}
\label{app:chambers}\label{app:highest-weight}
Choose $H_0\in\mathfrak t$ outside the root hyperplanes and put
$\Phi^+=\{\alpha:\alpha(H_0)>0\}$.
Define $C=\{H\in\mathfrak t:\alpha(H)>0\ \forall\alpha\in\Phi^+\}$;
its closure $\overline C$ uses weak inequalities.
For an empty root system both are $\mathfrak t$.

\begin{proposition}\label{prop:chamber}
Every $X\in\mathfrak g$ is conjugate into $\overline C$, and every regular
element into $C$. All maximal abelian subalgebras are conjugate.
\end{proposition}
These are the standard conjugacy and Weyl chamber results for compact
connected groups \cite{Varadarajan1984,ConradLandesman2018}.

For a finite-dimensional unitary representation, write
$V=\bigoplus_\mu V_\mu$, where $B_H|_{V_\mu}=\mu(H)I$ and $\mu\in P_T$.
The root convention gives
$d\rho(\mathfrak g_\alpha)V_\mu\subseteq V_{\mu+\alpha}$.

\begin{theorem}[Highest weight]\label{thm:highest-weight}
Every finite-dimensional irreducible representation has a unique highest
weight $\lambda$ with $V_\lambda=\C v_\lambda$.
Positive root operators annihilate $v_\lambda$, and every weight is
$\lambda$ minus a sum of positive roots with nonnegative integral
coefficients. Moreover
$2(\lambda,\alpha)/\|\alpha\|^2\in\Z_{\ge0}$ for every positive root.
For $H\in\overline C$, $\max\spec B_H=\lambda(H)$; for $H\in C$ the top
eigenspace is exactly $\C v_\lambda$.
\end{theorem}
The highest weight assertions are standard
\cite[Sections~3.1--3.2]{GoodmanWallach2009}.
For a general compact connected $G$, the center acts by a scalar character
on an irreducible block; the root theory applies to the semisimple part.
The spectral assertion follows directly: if
$\mu=\lambda-\sum_\alpha n_\alpha\alpha$, then
$\lambda(H)-\mu(H)=\sum_\alpha n_\alpha\alpha(H)\ge0$ on $\overline C$,
and it is strictly positive on $C$ unless $\mu=\lambda$.
Thus the top eigenspace assertion used in
Proposition~\ref{prop:stabilizer} needs no further argument.

\begin{theorem}[Casimir scalar]\label{thm:casimir}
In an irreducible block of highest weight $\lambda$,
\[
 \Omega=-\sum_jd\rho(X_j)^2=c_\lambda I,\qquad
 c_\lambda=\|\lambda\|^2+2(\rho_{\rm W},\lambda).
\]
\end{theorem}
\begin{proof}
The sum is independent of orthonormal basis, so invariance makes $\Omega$
commute with the group action and Schur's lemma makes it scalar.
At $v_\lambda$, its expectation is the squared norm of the generator mean
plus the total variance. Applying \eqref{eq:variance} and
\eqref{eq:total-variance}, and dividing by $\hbar^2$, gives the formula.
\end{proof}

\subsection{Technical details for the classical examples}\label{app:classical-details}
This subsection collects the harmonic decomposition proofs and normalization
checks used in Sections~\ref{sec:su}--\ref{sec:sp}.

\subsubsection{Real spherical harmonics}\label{app:classical-real}
\begin{lemma}\label{lem:spherical}
For real dimension $d\ge2$,
$-\Delta_S|_{\mathcal H_l}=l(l+d-2)I$ and
$L^2(S^{d-1})=\widehat\bigoplus_{l\ge0}\mathcal H_l$.
\end{lemma}
\begin{proof}
The polar formula
$\Delta=\partial_r^2+(d-1)r^{-1}\partial_r+r^{-2}\Delta_S$
applied to $r^lY$ gives the eigenvalue.
For harmonic homogeneous $h_s$,
\[
 \Delta(r^{2j}h_s)=2j(2s+d+2j-2)r^{2j-2}h_s .
\]
Decompose $\Delta p_l$ by induction on degree and lift each term with this
formula. Subtracting the lift gives
$p_l=\sum_{j=0}^{\lfloor l/2\rfloor}r^{2j}h_{l-2j}$.
Integration by parts makes $\Delta_S$ symmetric, so the different degree
spaces are orthogonal.

For $f\in C(S^{d-1})$, extend it continuously to a box by
$F(x)=\chi(|x|)f(x/|x|)$, with $\chi=0$ near zero and $\chi=1$ near the sphere.
Lemma~\ref{lem:bernstein} uniformly approximates this extension by polynomials.
The homogeneous decomposition shows their restrictions are finite sums of
spherical harmonics. Density of continuous functions in $L^2$ completes
the argument.
\end{proof}

\begin{lemma}\label{lem:so-irreducible}
For $n\ge3$, each $\mathcal H_l$ is irreducible under $SO(n)$, with highest
weight $l\varepsilon_1$ and normalized highest vector
$v_l=(x_1+ix_2)^l/\|(x_1+ix_2)^l\|_2$.
The block constants are those in \eqref{eq:so-constants}, with the same
formulas also valid for $l=0$.
\end{lemma}
\begin{proof}
Let $K=SO(n-1)$ fix the north pole.
Transitivity on perpendicular spheres makes every $K$-invariant homogeneous
polynomial a linear combination of
$x_n^{l-2j}(x_1^2+\cdots+x_{n-1}^2)^j$.
Odd perpendicular degrees vanish, since a rotation takes any perpendicular
vector to its negative.
Harmonicity gives the recurrence
\[
 2(j+1)(2j+n-1)a_{j+1}
       +(l-2j)(l-2j-1)a_j=0.
\]
Thus the fixed harmonic space is one-dimensional.
Every nonzero irreducible constituent has a vector nonzero at the north pole:
otherwise transitivity would make every function in it vanish everywhere.
Averaging such a vector over $K$ gives a nonzero fixed vector.
Two orthogonal constituents would give two independent fixed vectors.
Hence there is exactly one constituent.

Use the Cartan basis $E_{12},E_{34},\ldots$ and a regular element
$H_0=\sum_a b_aE_{2a-1,2a}$ with $b_1>\cdots>b_m>0$.
On degree-$l$ monomials, the unique maximal frequency is $lb_1$, attained
by $(x_1+ix_2)^l$.
The latter is harmonic, so it is a highest weight vector.
The dual coordinates are orthonormal.
The positive roots are $\varepsilon_i\pm\varepsilon_j$, $i<j$,
with $\varepsilon_i$ added in odd dimension.
Consequently
$\rho_{\rm W}=\frac12\sum_{a=1}^{\lfloor n/2\rfloor}(n-2a)\varepsilon_a$,
proving the norm and pairing claims.
The maximal pairing with a positive root is $l$, giving $C=l/2$.
\end{proof}

The covariance has entry $\hbar^2l/2$ in the $2(n-2)$ directions
$E_{1j},E_{2j}$, $j\ge3$, and zero in the remaining directions.
Indeed $D_{1j}(x_1+ix_2)^l=-lx_j(x_1+ix_2)^{l-1}$ and
$D_{2j}=iD_{1j}$ on this vector; parity eliminates the cross terms for distinct $j$.
The identity
$\sum_{i<j}D_{ij}^2=r^2\Delta-E^2-(n-2)E$, $E=\sum_i x_i\partial_i$,
then gives
\begin{equation}\label{eq:so-variance}
 \sum_{i<j}A_{ij}^2|_{\mathcal H_l}=\hbar^2l(l+n-2)I,\qquad
 V_{\rho(g)v_l}=\hbar^2l(n-2).
\end{equation}

\subsubsection{Complex spherical harmonics}\label{app:classical-complex}
Write $z_j=x_j+iy_j$,
$\partial_j=(\partial_{x_j}-i\partial_{y_j})/2$,
$\bar\partial_j=(\partial_{x_j}+i\partial_{y_j})/2$, and
$\Delta=4\sum_j\partial_j\bar\partial_j$.
Let $\mathcal P_{p,q}$ be the polynomials homogeneous of degree $p$ in $z$
and $q$ in $\bar z$, and let
$\mathcal H_{p,q}=(\ker\Delta\cap\mathcal P_{p,q})|_{S^{2n-1}}$.

\begin{lemma}\label{lem:complex-harmonics}
There is an orthogonal decomposition
$L^2(S^{2n-1})=\widehat\bigoplus_{p,q\ge0}\mathcal H_{p,q}$.
Each summand is irreducible under $SU(n)$, with highest weight
$\lambda_{p,q}=q\lambda_1+p\lambda_{n-1}$ and highest weight vector
\[
 v_{p,q}=z_n^p\bar z_1^q/\|z_n^p\bar z_1^q\|_2 .
\]
For $n\ge3$ these highest weights are pairwise distinct.
For $n=2$, all pairs with $p+q=l$ give the same spin $l/2$, with total
multiplicity $l+1$ in the sphere representation.
\end{lemma}
\begin{proof}
The real spherical harmonic decomposition in dimension $2n$
(Lemma~\ref{lem:spherical}) can be further decomposed under the scalar circle
$z\mapsto e^{it}z$. At total degree $l$, its distinct frequencies label the
pairs $p+q=l$, giving precisely the displayed orthogonal summands.
The Fischer decomposition
$\mathcal P_{p,q}=(\ker\Delta\cap\mathcal P_{p,q})
\oplus |z|^2\mathcal P_{p-1,q-1}$ gives
\begin{align*}
 \dim\mathcal H_{p,q}
 &=\binom{n+p-1}{p}\binom{n+q-1}{q}
   -\binom{n+p-2}{p-1}\binom{n+q-2}{q-1}\\
 &=\frac{p+q+n-1}{n-1}
       \binom{p+n-2}{n-2}\binom{q+n-2}{n-2}.
\end{align*}
Terms with a negative lower binomial index are zero.
The Fischer decomposition follows by applying $\Delta$ to powers of $|z|^2$
times a harmonic polynomial, and inducting on $\min(p,q)$.

The polynomial $w=z_n^p\bar z_1^q$ is harmonic since $n\ge2$.
For the pullback action,
$\rho(\exp tH)w=e^{it(qa_1-pa_n)}w$.
In the complexified action the positive matrix root $E_{ij}$, $i<j$, acts by
$E_{ij}z_k=-\delta_{ik}z_j$ and
$E_{ij}\bar z_k=\delta_{jk}\bar z_i$; it therefore annihilates $w$.
Thus $w$ generates a constituent of highest weight
$q\lambda_1+p\lambda_{n-1}$.
The Weyl dimension formula \cite[Chapter~7]{GoodmanWallach2009} at this weight is the last displayed expression,
so the constituent exhausts $\mathcal H_{p,q}$.
For $n\ge3$, $\lambda_1,\lambda_{n-1}$ are independent; for $n=2$ they
coincide and $\dim\mathcal H_{p,q}=p+q+1$.
Counting the $l+1$ pairs with $p+q=l$ proves the multiplicity assertion.
\end{proof}

\subsubsection{Fundamental normalization in type A}
\label{subsec:su-fundamental}
The natural action on $\C^n$ is irreducible: transitivity on the unit sphere
implies that the orbit of any nonzero vector spans $\C^n$.
For the metric \eqref{eq:su-metric}, $e_1$ has highest weight $\lambda_1$.
The positive root matrices $E_{ij}$, $i<j$, annihilate $e_1$; their unit
normalizations in the complexified Hermitian metric are $E_{ij}/\sqrt2$.
Their brackets with $H$ are $[H,E_{ij}]=i(a_i-a_j)E_{ij}$.
Transitivity on rays and Proposition~\ref{prop:stabilizer} give
\[
 \Orb_{\hbar\lambda_1}\cong\mathbb{CP}^{n-1},\qquad
 |q|=\hbar\sqrt{\frac{n-1}{2n}},\qquad
 V_{\psi_g}=\frac{\hbar^2(n-1)}2.
\]
Hence the sufficient conditions for inclusion are
\begin{equation}\label{eq:su-fundamental}
 \hbar\sqrt{\frac{n-1}{2n}}\le M-\delta_\eta,\qquad
 \hbar^2\le\frac{\eta^3}{4n^2(n-1)}.
\end{equation}
This finite-dimensional computation checks the normalization.
In this range $\|A_j\|\le\hbar/\sqrt2\le\eta/8$ and the orbit radius is
at most $\eta/8$, so the qualified ball $B(0,\eta)$ already covers the orbit.
The growing blocks $\mathcal H_{0,k}$ above provide the fixed nonzero orbit
at resolutions tending to zero.

\subsubsection{Quaternionic spherical harmonics}\label{app:classical-quaternionic}
\begin{lemma}\label{lem:quaternion-harmonics}
Let $n\ge2$ and let $\mathcal H_l$ be the degree-$l$ spherical harmonics in
real dimension $4n$. The decomposition in \eqref{eq:quaternion-decomposition}
is a decomposition under $Sp(n)\times Sp(1)$,
where $\mathcal V_{(a,b)}$ is irreducible of highest weight
$\lambda_{a,b}=a\varepsilon_1+b\varepsilon_2$.
Under $Sp(n)$ alone its multiplicity is $a-b+1$.
For $n=1$ use the regular-representation decomposition in Section~\ref{sec:sp}.
\end{lemma}
\begin{proof}
The quaternionic harmonic decomposition and dimensions are given in
\cite[Propositions 2.1 and 3.1]{Ahrens2020}; its indices are
$h=a+b$, $m=b$, and its right $Sp(1)$ weight is $h-2m=a-b$.
Quaternionic conjugation converts the left scalar convention there to the
right scalar convention here.

For completeness, identify the $Sp(n)$ highest weight in each component.
The complexified linear coordinate space is $V\otimes\C^2$, with $V$ the
standard symplectic module. Denote its two columns by $x_i,y_i$,
$1\le i\le2n$, with the first $n$ entries of weights $+\varepsilon_i$.
The polynomial
$x_1^{a-b}(x_1y_2-x_2y_1)^b$
has $Sp(n)\times Sp(1)$ highest weight $(a\varepsilon_1+b\varepsilon_2,a-b)$.
The positive root matrices annihilate it. It is harmonic since Laplacian
contractions pair opposite $Sp(n)$ weights, whereas only positive-weight
variables occur.
It therefore generates the corresponding irreducible constituent.

The Weyl dimension formula for $\mathcal V_{(a,b)}$ is
\[
 d_{a,b}=
 \frac{(a-b+1)(a+b+2n-1)}{(2n-2)(2n-1)}
 \binom{a+2n-2}{2n-3}\binom{b+2n-3}{2n-3}.
\]
The joint harmonic component has dimension $(a-b+1)d_{a,b}$ by the
cited dimension formula, so this constituent exhausts it.
Restriction to $Sp(n)$ leaves the second tensor factor as a multiplicity
space. Choose an orthonormal basis in it to obtain individual irreducible
blocks for Assumption~\ref{asm:blocks}.
\end{proof}
In particular, $\dim\mathcal H_1=4n=2\dim\C^{2n}$.
At degree two the dimensions are
$3n(2n+1)+n(2n-1)-1=8n^2+2n-1$.

\subsubsection{Fundamental normalization in type C}\label{app:sp-fundamental}
This subsection also applies when $n=1$.
The real algebra and its complexification are
\[
 \mathfrak{sp}(n)=
 \left\{\begin{pmatrix}A&B\\-\bar B&\bar A\end{pmatrix}:
 A^*=-A,\ B^{\mathsf T}=B\right\},\qquad
 \mathfrak{sp}(n)_\C=
 \left\{\begin{pmatrix}A&B\\C&-A^{\mathsf T}\end{pmatrix}:
 B^{\mathsf T}=B,\ C^{\mathsf T}=C\right\}.
\]
Quaternionic Gram--Schmidt shows transitivity on the unit sphere, proving
irreducibility of the natural complex representation on $\C^{2n}$.
Its highest vector is $e_1$ and its highest weight is $\lambda_1=\varepsilon_1$.
Positive root matrices can be chosen as
\[
 Z_{\varepsilon_i-\varepsilon_j}=
 \begin{pmatrix}E_{ij}&0\\0&-E_{ji}\end{pmatrix},\quad
 Z_{\varepsilon_i+\varepsilon_j}=
 \begin{pmatrix}0&E_{ij}+E_{ji}\\0&0\end{pmatrix},\quad
 Z_{2\varepsilon_i}=\begin{pmatrix}0&E_{ii}\\0&0\end{pmatrix}.
\]
For the first two families we take $i<j$.
Their brackets with $H$ are $i\alpha(H)Z_\alpha$.
Their first columns vanish, so they annihilate $e_1$.
Compact conjugation is $Z\mapsto-Z^*$; the complexified Hermitian norm is
$2\tr(Z^*Z)$.
The first two families have squared norm $4$ and the last has squared norm $2$;
the unit vectors are consequently $Z_\alpha/2$ and $Z_{2\varepsilon_i}/\sqrt2$.

Transitivity on complex rays gives
\[
 \Orb_{\hbar\lambda_1}\cong\mathbb{CP}^{2n-1},\qquad
 |q|=\hbar/2,\qquad
 (\rho_{\rm W},\lambda_1)=n/4,\qquad V_{\psi_g}=\hbar^2n/2.
\]
The ray stabilizer is $U(1)\times Sp(n-1)$, of real dimension
$2n^2-3n+2$; its quotient has dimension $4n-2$.
Sufficient conditions for the fundamental orbit inclusion are
\begin{equation}\label{eq:sp-fundamental}
 \hbar/2\le M-\delta_\eta,\qquad
 \hbar^2\le\frac{\eta^3}{4n(2n^2+n+1)}.
\end{equation}
As in the $SU(n)$ fundamental computation, these conditions imply
$\|A_j\|\le\hbar/\sqrt2\le\eta/8$, so one qualified zero-centered ball covers
this small orbit. The growing function-space blocks provide the corresponding
fixed nonzero orbit.

\subsubsection{Polynomial orbit equations}\label{app:classical-orbit-equations}
\begin{theorem}[Invariant polynomials separate compact orbits]\label{thm:orbit-polynomials}
For a compact group acting orthogonally on $\R^d$, let
$\R[\R^d]^G=\{p:p(gx)=p(x)\ \forall g,x\}$.
Then for every $q$,
\[
 G.q=\{x:p(x)=p(q)\ \text{for every }p\in\R[\R^d]^G\}.
\]
\end{theorem}
\begin{proof}
If $x\notin G.q$, set $d_0=\dist(x,G.q)>0$ and
$F(y)=\dist(y,G.q)^2$.
This is continuous and invariant, with $F(q)=0$, $F(x)=d_0^2$.
On a box containing an invariant ball containing both orbits, approximate
$F$ by a real polynomial $Q$ with error less than $d_0^2/3$, using
Lemma~\ref{lem:bernstein}.
The Haar average $p(y)=\int_GQ(gy)\,dg$ is an invariant polynomial of bounded
degree and retains that error on the ball.
Thus $p(x)>2d_0^2/3$ and $|p(q)|<d_0^2/3$, separating the two points.
\end{proof}
The theorem concerns the common real level set of all invariants.
For the finite equations used next, sufficiency follows from a separate
normal-form calculation.

\begin{proof}[Proof of Proposition~\ref{prop:plucker}]
Associate to $q$ its real skew-symmetric matrix $Q$.
For $Q\ne0$, choose $|u|=1$ with $-Q^2u=r^2u$, $r>0$, and put $v=-Qu/r$.
Then $u\perp v$, $|v|=1$, $Qu=-rv$, and $Qv=ru$.
The resulting plane and its perpendicular complement are $Q$-invariant.
Induction gives an orthonormal basis in which
$q=\sum_a r_ae_{2a-1}\wedge e_{2a}$.
Hence
\[
 q\wedge q=2\sum_{a<b}r_ar_b
 e_{2a-1}\wedge e_{2a}\wedge e_{2b-1}\wedge e_{2b}.
\]
It vanishes exactly when at most one coefficient is nonzero.
Expanding in arbitrary coordinates gives the displayed equations.
Normalization and transitivity of oriented two-frames prove the orbit claim.
\end{proof}
The scalar invariant interpretation is described after Proposition~\ref{prop:plucker}.

\subsubsection{The truncated-cone estimate}\label{app:cone-estimate}
\begin{proof}[Proof of Theorem~\ref{thm:cone}]
Fix $y=ru\wedge v$, $0\le r\le M$, and any center $\xi$ with
$|\xi-y|\le\delta_\eta$. Independently choose $l\ge0$ with
$|\hbar l-r|\le\hbar/2$ and a coherent state with mean
$m=\hbar l\,u\wedge v$.
For this center one has
\[
 V_\psi\le\hbar(M+\hbar/2)(n-2),\qquad
 |m-\xi|^2\le\hbar^2/2+2\delta_\eta^2.
\]
Since $\hbar<1$, $n\ge3$, and $M\ge1$,
\begin{align*}
 V_\psi+|m-\xi|^2
 &\le2nM\hbar+2\delta_\eta^2\\
 &\le\frac{\eta^3}{4(N+1)}+\frac{\eta^4}{512N^2}
 \le\frac{9\eta^3}{32(N+1)}.
\end{align*}
The last inequality follows from $\eta(N+1)\le16N^2$.
The offset estimate \eqref{eq:offset} gives a loss at most $\eta/4$.
For the default grid, first choose $\xi\in D_\eta$ near $y$;
then $|\xi|\le|y|+\delta_\eta\le M+\delta_\eta$.
For the regional grid choose $\xi\in E_{\eta,M}$ using its covering property;
by construction $|\xi|\le M$. In both cases
$|y-\xi|\le\delta_\eta<\eta$ proves inclusion.
If $l=0$, the constant angular state has zero variance and the same bounds apply.
On $L^2(\R^n)$ tensor with a unit radial vector; the angular action gives the
same common generator domain and cutoff norms.
\end{proof}

\ifssphandbook
\section{Quick Reference for Readers}
This optional appendix gives a short route through the objects used in the
paper. Appendix~\ref{app:prelim} records the standard results and their
sources, together with the domain arguments and classical computations.

\subsection{Hilbert spaces and operator domains}
A Hilbert space is a complete inner product space; convergence is measured by
the norm $\|v\|=\sqrt{\langle v,v\rangle}$.
A bounded operator acts continuously on the whole space.
An unbounded observable is specified by both its formula and its dense domain.
For example, multiplication by $x$ on $L^2(\R)$ has domain
$\{f:xf\in L^2\}$.
Self-adjointness includes equality with the adjoint domain, and a core is a
smaller domain dense in the graph norm $\|v\|_A$.

\subsection{Spectra, spectral measures, and cutoff functions}
In finite dimension, a self-adjoint matrix is a weighted sum of its
orthogonal eigenprojections.
The spectral theorem replaces that sum by $A=\int t\,dE_A(t)$.
For a unit state $\psi$, $\mu_\psi(S)=\langle\psi,E_A(S)\psi\rangle$ is the
probability of observing a value in $S$.
Functional calculus gives
\[
 \langle\psi,f(A)\psi\rangle=\int f\,d\mu_\psi,\qquad
 \|f(A)\psi\|^2=\int|f|^2\,d\mu_\psi,\qquad
 f\ge0\Longrightarrow f(A)\ge0 .
\]
Our cutoff equals one near its center and zero outside a slightly larger
interval. Its application to $A$ is a bounded positive contraction.

\subsection{Stone's theorem}
A strongly continuous one-parameter unitary group $U(t)$ determines a
self-adjoint generator $B$
through $U(t)=e^{itB}$.
The derivative exists exactly on $\Dom B$, and there
$Bv=\lim_{t\to0}(U(t)v-v)/(it)$.
For a Lie group representation, apply this separately to each curve
$\rho(\exp tX)$.
The physical scaling in this paper is $A_X=\hbar B_X$.

\subsection{Lie groups and coadjoint orbits}
A Lie group combines smooth geometry with multiplication.
Its Lie algebra consists of tangent vectors at the identity with their bracket.
Conjugation acts linearly on this algebra; the dual action transports linear
functionals and their coordinate vectors of expectations.
The coadjoint orbit of $\ell$ is the set of all transported functionals.
For $SO(3)$ these are spheres; for a Heisenberg group at nonzero central
coordinate they are affine planes.

\subsection{Weights, roots, and highest weight vectors}
A maximal torus is a maximal connected compact abelian subgroup.
Its commuting generators can be diagonalized, and their simultaneous
eigenvalue functionals are weights.
Roots describe how the algebra moves one weight space to another.
After choosing positive roots, a highest weight vector is an eigenvector
annihilated by every raising operator.
In a finite-dimensional irreducible compact-group representation its line
is unique, and translating that line gives the coherent-state ray orbit.

\subsection{Harmonics and projective quotients}
Spherical harmonics are restrictions of homogeneous harmonic polynomials.
They decompose square-integrable functions on a sphere into finite-dimensional
angular pieces.
Complex harmonics refine the degree to a pair $(p,q)$.
Quaternionic harmonics separate the commuting $Sp(n)$ and $Sp(1)$ actions,
and their multiplicity under $Sp(n)$ is recorded in
\eqref{eq:quaternion-decomposition}.
Passing to a projective quotient selects fixed functions of the scalar
group $O(1)$, $U(1)$, or $Sp(1)$.

\subsection{A complete spin-\texorpdfstring{$1/2$}{1/2} calculation}
Take $J_a=\hbar\sigma_a/2$ on $\C^2$, $\psi=(1,0)^{\mathsf T}$, and
$0<\eta\le1/2$. Then
\begin{align*}
 m_\psi&=(0,0,\hbar/2),\\
 \Var_\psi(J_x)=\Var_\psi(J_y)&=\hbar^2/4,\qquad\Var_\psi(J_z)=0,\\
 V_\psi&=\hbar^2/2,\\
 \hbar^2\le\eta^3/16&\Longrightarrow V_\psi\le\eta^3/32,\\
 \cut_{0,\eta}(J_a)=I&\Longrightarrow
 S^2_{\hbar/2}\subseteq B(0,\eta)\subseteq\sSp^\eta_{D_\eta,M}(J).
\end{align*}
For the last line the same bound gives
$\|J_a\|=\hbar/2\le\eta/8$, so the zero-centered factors are indeed $I$.
Rotating $\psi$ also constructs qualified centers around every point of its
mean sphere.

\subsection{Spin representations and descent to \texorpdfstring{$SO(3)$}{SO(3)}}
\label{subsec:spin}
Let $\sigma_x,\sigma_y,\sigma_z$ be the Pauli matrices and put
$E_a=i\sigma_a/2$. They are orthonormal for $-2\tr(XY)$ and satisfy
$[E_x,E_y]=-E_z$ and its cyclic variants.
For $j\in\frac12\Z_{\ge0}$, the symmetric tensor construction
\cite{Woit2017} gives
\[
 V_j=\operatorname{Sym}^{2j}\C^2,\qquad
 \rho_j(U)=U^{\otimes 2j}|_{V_j},\qquad J_a=-i\hbar\,d\rho_j(E_a).
\]
The normalized symmetric tensor basis $|s\rangle$, $s=-j,\ldots,j$, obeys
\begin{equation}\label{eq:ladder}
 J_z|s\rangle=\hbar s|s\rangle,\qquad
 J_+|s\rangle=\hbar\sqrt{(j-s)(j+s+1)}\,|s+1\rangle,\quad J_-=J_+^*.
\end{equation}
To obtain the coefficient, put $r=j+s$. A tensor in $|s\rangle$ has
$r$ factors of weight $+1/2$ and $2j-r$ of weight $-1/2$. Raising one
negative-weight factor produces each resulting tensor $r+1$ times;
normalization therefore gives
$(r+1)\sqrt{\binom{2j}{r+1}/\binom{2j}{r}}$.
The nonzero ladder coefficients and the spectral projections of $J_z$
prove irreducibility. The highest vector is $|j\rangle$.
Direct computation yields
\[
 \sum_aJ_a^2=\hbar^2j(j+1)I,\quad
 m_{|j\rangle}=(0,0,\hbar j),\quad
 \Cov_{|j\rangle}=\operatorname{diag}(\hbar^2j/2,\hbar^2j/2,0).
\]
In the Cartan coordinate dual to $E_z$, the positive root has length one,
$\lambda_j=j\varepsilon$, and $\rho_{\rm W}=\varepsilon/2$.
This coordinate $\varepsilon$ is local to the spin convention.

Conjugation of traceless Hermitian matrices by $SU(2)$ gives a surjective
homomorphism onto $SO(3)$ with kernel $\{\pm I\}$.
Its derivative is an isomorphism, so its compact image is open and closed
in the connected rotation group.
Since $\rho_j(-I)=(-1)^{2j}I$, this representation descends to $SO(3)$
exactly for integral $j$.
For every allowed $SU(2)$ spin the coadjoint orbit is
$S^2_{\hbar j}=\{x\in\R^3:|x|=\hbar j\}$.

\subsection{A sparse spin direct sum}\label{app:sparse-spin}
This example illustrates Theorem~\ref{thm:direct-sum}.
Define
\[
 K_3(G,M)=\max\{1,16(N+1)M\|\rho_{\rm W}\|\}.
\]
The condition $\hbar\le\eta^3/K_3(G,M)$ ensures the common variance bound
for every radius-selected block.
For the spin convention, $K_3(SU(2),M)=\max\{1,32M\}$.

\begin{proposition}\label{prop:sparse}
Fix $\rho_{\rm sp}=\widehat\bigoplus_{\nu\ge0}V_{2^\nu}$, $M=11/10$,
and $K_3=176/5$. For $a\in\{1,3/2\}$ set
$\hbar_k^{(a)}=a2^{-k}$, $\eta_k^{(a)}=(K_3\hbar_k^{(a)})^{1/3}$,
and $S_k^{(a)}=\sSp^{\eta_k^{(a)}}_{D_{\eta_k^{(a)}},M}(A^{\hbar_k^{(a)}})$.
For every $k\ge9$,
\[
 F_1=\{0\}\cup\bigcup_{m\ge0}S^2_{2^{-m}}\subseteq S_k^{(1)},\qquad
 F_{3/2}=\{0\}\cup\bigcup_{m\ge0}S^2_{(3/4)2^{-m}}\subseteq S_k^{(3/2)}.
\]
The target sets are distinct.
\end{proposition}
\begin{proof}
Write $\hbar=\hbar_k^{(a)}$ and $\eta=\eta_k^{(a)}$.
For $k\ge9$ one has $\eta\le1/2$ and
$\delta_\eta\le1/384<M-1$.
The block with spin $2^\nu$ has radius $r_\nu=a2^{\nu-k}$ and
$\hbar^2(\rho_{\rm W},\lambda_\nu)=\hbar r_\nu/2$.
Radius selection gives $\nu=0,\ldots,k$ when $a=1$, and
$\nu=0,\ldots,k-1$ when $a=3/2$.
The largest selected radii are $1$ and $3/4$.
The variance condition is $32r_\nu\le K_3$, valid in both cases.
Theorem~\ref{thm:direct-sum} therefore includes every selected sphere.

The remaining target spheres have radius less than $\hbar$.
Use the actual smallest block, spin $1$: each generator has spectrum in
$[-\hbar,\hbar]$. Since $\hbar/\eta=\eta^2/K_3<3/4$,
all zero-centered cutoff factors are $I$ on this block.
Thus $B(0,\eta)\subseteq S_k^{(a)}$, which covers the remaining spheres and
the origin. Finally $F_1$ contains a unit vector and every point of $F_{3/2}$
has norm at most $3/4$.
\end{proof}
These are two explicit target inclusions for two parameter sequences of the
same representation, in the sense of Remark~\ref{rem:scope}.
\fi

\section{A Noncompact Comparison: the \texorpdfstring{$(2n+1)$}{(2n+1)}-Dimensional
Heisenberg Group}\label{app:heisenberg}
This comparison uses the self-adjoint cutoff tools in
Section~\ref{sec:cutoff} with Gaussian states.
It separates the dimension of the group from the length of the physical tuple.
The group is two-step nilpotent and is represented by matrices of size $n+2$.

\subsection{The group and its self-adjoint generators}
For $n\ge1$ define
\[
 H_n=\left\{\begin{pmatrix}1&u^{\mathsf T}&w\\0&I_n&v\\0&0^{\mathsf T}&1\end{pmatrix}:
 u,v\in\R^n,\ w\in\R\right\}.
\]
Its multiplication is
$(u,v,w)(u',v',w')=(u+u',v+v',w+w'+u\cdot v')$.
The algebra basis
$Q_i=E_{1,i+1}$, $P_j=E_{j+1,n+2}$, $C=E_{1,n+2}$
has brackets $[Q_i,P_j]=\delta_{ij}C$ and all other basis brackets zero.
In dual coordinates $(a,b,c)$,
\[
 \Ad^*(\exp(u\cdot Q+v\cdot P))(a,b,c)
       =(a+cv,b-cu,c).
\]
Indeed $\Ad(\exp(-u\cdot Q-v\cdot P))Q_i=Q_i+v_iC$ and
the corresponding formula for $P_j$ is $P_j-u_jC$.
For $c\ne0$ an orbit is the whole affine plane
$\R^{2n}\times\{c\}\subseteq\R^{2n+1}$; for $c=0$ it is a point.

On $L^2(\R^n,dx)$ put
\[
 \rho_\hbar(u,v,w)f(x)=e^{i(w+v\cdot x)/\hbar}f(x+u).
\]
Composing two such operators gives the required cross phase $u\cdot v'/\hbar$.
Translations and multiplication by modulus-one functions preserve the norm.
Continuity on $C_c^\infty$ and density give strong continuity on $L^2$.
Differentiation gives
\[
 d\rho_\hbar(Q_i)=\partial_i,\qquad
 d\rho_\hbar(P_j)=ix_j/\hbar,\qquad
 d\rho_\hbar(C)=i/\hbar .
\]
Thus the complete tuple of group generators includes
$A_{Q_i}=P_{\hbar,i}$, $A_{P_j}=X_j$, and $A_C=I$, where
\begin{align*}
 X_jf&=x_jf,& \Dom X_j&=\{f\in L^2:x_jf\in L^2\},\\
 P_{\hbar,j}f&=-i\hbar\partial_jf,&
 \Dom P_{\hbar,j}&=\{f\in L^2:\partial_j f\in L^2\}.
\end{align*}
The derivatives are weak derivatives:
$\int f\,\partial_j\varphi=-\int(\partial_j f)\varphi$ for every
$\varphi\in C_c^\infty$.
Their domain intersection over all momentum directions is $H^1(\R^n)$.
Multiplication is self-adjoint on the stated domain, by testing its adjoint
on compactly supported functions.
The Fourier transform identifies $P_{\hbar,j}$ with the real multiplication
operator $\hbar M_{\xi_j}$, proving self-adjointness.

The Schwartz space is
\[
 \mathcal S(\R^n)=\{f\in C^\infty:
 \sup_x|x^\alpha\partial^\beta f(x)|<\infty
 \text{ for all }\alpha,\beta\in\Z_{\ge0}^n\}.
\]
It is invariant under all these generators and is a common core.
Here is the full graph-norm approximation.
Choose $\chi\in C_c^\infty$ equal to one near the unit ball and
$\chi_R(x)=\chi(x/R)$, with $0\le\chi\le1$.
For $f\in\Dom X_j$, dominated convergence gives
$\chi_Rf\to f$ and $x_j\chi_Rf\to x_jf$ in $L^2$.
For $f\in\Dom P_{\hbar,j}$,
\[
 \partial_j(\chi_Rf)=\chi_R\partial_j f+(\partial_j\chi_R)f,\qquad
 \|(\partial_j\chi_R)f\|_2\le R^{-1}\|\partial_j\chi\|_\infty\|f\|_2.
\]
Thus cutoff approximation converges in each full natural graph norm.

Choose a smooth nonnegative kernel $\kappa$ supported in the unit ball,
with integral one, and set $\kappa_\varepsilon(x)=\varepsilon^{-n}
\kappa(x/\varepsilon)$.
For fixed $R$, put $u_R=\chi_Rf$ and
$g_{R,\varepsilon}=u_R*\kappa_\varepsilon\in C_c^\infty$.
Strong continuity of translations gives
$w*\kappa_\varepsilon\to w$ in $L^2$, and
$\|w*k\|_2\le\|w\|_2\|k\|_1$.
For position,
\[
 x_jg_{R,\varepsilon}
 =(x_ju_R)*\kappa_\varepsilon+
                  u_R*(x_j\kappa_\varepsilon),\qquad
 \|x_j\kappa_\varepsilon\|_1\le\varepsilon .
\]
For momentum,
$\partial_jg_{R,\varepsilon}=(\partial_j u_R)*\kappa_\varepsilon$.
Both graph norms therefore converge to those of $u_R$ as
$\varepsilon\to0$.
Choose a diagonal sequence $R\to\infty$, $\varepsilon_R\to0$ with each graph
error tending to zero.
For a vector in the common domain choose the same $\varepsilon_R$ for all
$2n$ errors. This proves the common core assertion on the full domains.

\subsection{Gaussian states and the physical phase-space spectrum}
The physical tuple used for the synthetic spectrum is
\begin{equation}\label{eq:physical-tuple}
 A_\parallel=(X_1,\ldots,X_n,P_{\hbar,1},\ldots,P_{\hbar,n}),\qquad N=2n.
\end{equation}
For $(q,p)\in\R^{2n}$ let
\[
 \psi_{q,p}(x)=(\pi\hbar)^{-n/4}
     e^{-|x-q|^2/(2\hbar)}e^{ip\cdot x/\hbar}\in\mathcal S(\R^n).
\]
The one-dimensional integrals
$\int e^{-y^2/\hbar}dy=\sqrt{\pi\hbar}$,
$\int ye^{-y^2/\hbar}dy=0$, and
$\int y^2e^{-y^2/\hbar}dy=(\hbar/2)\sqrt{\pi\hbar}$
factor across coordinates.
Since $P_{\hbar,j}\psi_{q,p}=(p_j+i(x_j-q_j))\psi_{q,p}$, they give
\begin{equation}\label{eq:gaussian-variance}
 \|\psi_{q,p}\|=1,\quad m_{\psi_{q,p}}=(q,p),\quad
 \Var(X_j)=\Var(P_{\hbar,j})=\hbar/2,\quad V_{\psi_{q,p}}=n\hbar .
\end{equation}
The real covariance is $(\hbar/2)I_{2n}$.
For the same coordinate the mixed complex pairing equals $i\hbar/2$,
whose real part is zero.
The complete group expectation, in $(Q,P,C)$ order, is $(p,q,1)$ on the
central coadjoint plane.

\begin{proposition}\label{prop:heisenberg}
Let $M>0$ and $0<\eta\le1/2$. If
\begin{equation}\label{eq:heisenberg-threshold}
 0<\hbar\le\frac{\eta^3}{8n(2n+1)},
\end{equation}
the grids defined with $N=2n$ satisfy
\[
 \sSp^\eta_{D_\eta}(A_\parallel)=\R^{2n},\qquad
 B(0,M)\subseteq\sSp^\eta_{E_{\eta,M},M}(A_\parallel)
                         \subseteq B(0,M+\eta).
\]
Also $B(0,M-\delta_\eta)\subseteq\sSp^\eta_{D_\eta,M}(A_\parallel)$
whenever $M\ge\delta_\eta$.
\end{proposition}
\begin{proof}
Given any center $\xi=(q,p)$, use its Gaussian in \eqref{eq:gaussian-variance}.
Lemma~\ref{lem:tail} and Lemma~\ref{lem:product} give
\begin{equation}\label{eq:heisenberg-loss}
 \|\Cuts_{\xi,\eta}(A_\parallel)\|
 \ge1-\frac{2n+1}{2}\frac{16n\hbar}{\eta^2}
 =1-\frac{8n(2n+1)\hbar}{\eta^2}\ge1-\eta .
\end{equation}
Every physical center is qualified.
The default covering radius is less than $\eta$ and the regional grid covers
the full closed ball, proving the assertions.
The regional outer bound follows directly from its center restriction.
\end{proof}
For $n=1$ the coefficient $(2n+1)/2$ is $3/2$, giving the loss
$24\hbar/\eta^2$ from Proposition~\ref{prop:position-momentum}.

Including the center generator gives a different tuple:
its central cutoff is $\cut_{\xi_C,\eta}(I)=\cut_{\xi_C,\eta}(1)I$.
For $|\xi_C-1|\ge\eta$ that factor is zero, so all qualified centers have
$|\xi_C-1|<\eta$.
The associated ball union is contained in $|x_C-1|<2\eta$.
This explains why the full phase-space conclusion above uses precisely the
$2n$ physical coordinates in \eqref{eq:physical-tuple}.

\printbibliography[title={References}]

\clearpage
\section*{Author addresses}
\noindent\textsc{Junyang Liu}\\
School of Mathematical Sciences, East China Normal University,
Shanghai, China\\
\textit{E-mail address:}
\href{mailto:10231510429@stu.ecnu.edu.cn}{\texttt{10231510429@stu.ecnu.edu.cn}}

\par\medskip
\noindent\textsc{Hang Wang}\\
Research Center for Operator Algebras, East China Normal University,
Shanghai 200062, China\\
\textit{E-mail address:}
\href{mailto:wanghang@math.ecnu.edu.cn}{\texttt{wanghang@math.ecnu.edu.cn}}

\par\medskip
\noindent\textsc{Houjin Wang}\\
School of Mathematical Sciences, East China Normal University,
Shanghai, China\\
\textit{E-mail address:}
\href{mailto:10222140426@stu.ecnu.edu.cn}{\texttt{10222140426@stu.ecnu.edu.cn}}

\par\medskip
\noindent\textsc{Haoran Xu}\\
School of Mathematical Sciences, East China Normal University,
Shanghai, China\\
\textit{E-mail address:}
\href{mailto:10231510419@stu.ecnu.edu.cn}{\texttt{10231510419@stu.ecnu.edu.cn}}
\end{document}